\documentclass[paper]{JHEP}
\usepackage{epsfig}

\def\gsim{\ \rlap{\raise 3pt \hbox{$>$}}{\lower 3pt \hbox{$\sim$}}\ }
\def\lsim{\ \rlap{\raise 3pt \hbox{$<$}}{\lower 3pt \hbox{$\sim$}}\ }
\title{Trojan Penguins and Isospin Violation in Hadronic B Decays}

\author{Yuval Grossman and Matthias Neubert\thanks{On leave from: 
Newman Laboratory of Nuclear Studies, Cornell University, Ithaca, 
NY 14853}\\
Stanford Linear Accelerator Center, Stanford University\\
Stanford, California 94309, U.S.A.\\
E-mail: \email{yuval@slac.stanford.edu}, 
\email{neubert@slac.stanford.edu}}

\author{Alexander L. Kagan\\
Department of Physics, University of Cincinnati\\
Cincinnati, Ohio 45221, U.S.A.\\
E-mail: \email{kagan@physics.uc.edu}}

\abstract{Some rare hadronic decays of $B$ mesons, such as 
$B\to\pi K$, are sensitive to isospin-violating contributions
from physics beyond the Standard Model. Although commonly referred 
to as electroweak penguins, such contributions can often arise 
through tree-level exchanges of heavy particles, or through 
strong-interaction loop diagrams. The Wilson coefficients of the 
corresponding electroweak penguin operators are calculated in a 
large class of New Physics models, and in many cases are found
not to be suppressed with respect to the QCD penguin coefficients.
Several tests for these effects using observables in 
$B^\pm\to\pi K$ decays are discussed, and nontrivial bounds on the 
couplings of the various New Physics models are derived.}


\preprint{SLAC-PUB-8243\\
\hepph{9909297}}

\begin{document}

\section{Introduction}

The study of rare decay processes is an important tool in 
testing the fundamental interactions among elementary particles, 
exploring the origin of CP violation, and searching for New 
Physics beyond the Standard Model (SM). Such processes have been 
explored in great detail, both theoretically and in experimental 
searches, in the weak interactions of kaons and $B$ mesons, as 
well as in $K$--$\bar K$, $D$--$\bar D$, and $B$--$\bar B$ mixing. 
Two prominent examples are the evidence for mixing-induced 
CP violating in the decay $B\to J/\psi\,K_S$ reported by the CDF 
Collaboration \cite{CDF}, and the observation of direct CP 
violation in the decays $K\to\pi\pi$ \cite{NA31,E731}, which has 
recently been confirmed by the KTeV and NA48 Collaborations 
\cite{KTeV,NA48}. 

Flavor-changing neutral current (FCNC) processes, which are 
forbidden at tree level in the SM, are especially sensitive to 
any new source of flavor-violating interactions. Already, the 
absence of experimental signals for new FCNC couplings puts 
stringent bounds on the parameters of many 
extensions of the SM such as Supersymmetry (SUSY) \cite{Yuval}.
So far, FCNC processes have been explored mainly in 
particle--antiparticle mixing and in ``semihadronic'' weak decays,
which permit a clean theoretical description. In the kaon system, 
examples of the latter type are the decays $K\to\pi\,l^+l^-$ and 
$K\to\pi\,\nu\bar\nu$. In the $B$ system, the decays that have 
received the most attention are $B\to X_s\,\gamma$, 
$B\to X_s\,l^+l^-$ and $B\to X_s\,\nu\bar\nu$, where $X_s$ can be 
any final state, exclusive or inclusive, containing a strange 
quark \cite{BaBar}. 

In the present paper, we explore in detail how New Physics could 
affect purely hadronic decays such as $B\to\pi K$, which are 
sensitive to isospin- or, more generally, SU(3) flavor-violating 
interactions. In the SM, the main contributions to the decay 
amplitudes for these processes come from the penguin-induced FCNC 
transition $\bar b\to\bar s q\bar q$, which by far exceeds a small, 
Cabibbo-suppressed $\bar b\to\bar u u\bar s$ contribution from 
$W$-boson exchange. Because of a fortunate interplay of isospin, 
Fierz and flavor symmetries, the theoretical description of the 
charged decays $B^\pm\to\pi K$ in the SM is clean despite 
the fact that these are exclusive nonleptonic decays \cite{NR,me}. 
Isospin violation arises through the small charged-current 
contribution and through the electroweak penguin operators in the 
low-energy effective weak Hamiltonian \cite{Heff}. In the SM, 
these operators are induced by penguin and box diagrams involving 
the exchange of weakly interacting $W$ and $Z$ bosons, or of a 
photon. Here we point out that in a large class of New Physics 
models such effects can be mediated by ``trojan'' electroweak 
penguins, which are neither (pure) penguins nor of electroweak 
origin. Nevertheless, at low energies their effects are 
parameterized by an extension of the usual basis of electroweak 
penguin operators. We will explore examples where trojan penguins 
are induced by tree-level couplings, e.g., models with an extra 
$Z'$ boson, or pure strong-interaction processes, e.g., gluino box 
diagrams in SUSY models. 

In the present work we calculate the Wilson coefficients of the 
hadronic electroweak penguin operators in the effective weak 
Hamiltonian in an extended operator basis and for a large class 
of New Physics models. We then explore the phenomenological 
consequences of these new, isospin-violating contributions for 
weak-interaction observables. This serves two purposes: first, 
it allows us to derive bounds on New Physics parameters, which in 
some cases improve upon existing bounds derived from other 
processes; secondly, it shows which observables may be interesting 
to look at as far as searches for New Physics are concerned. We 
shall address both issues in detail for the particular case of 
$B^\pm\to\pi K$ decays, which in the context of the SM are of 
prime importance in determining the weak phase 
$\gamma=\mbox{arg}(V_{ub}^*)$ of the Cabibbo--Kobayashi--Maskawa 
(CKM) matrix \cite{NR,me,FM,Robert,mat,Frank}. We stress, however, 
that trojan penguins could be important in a much wider class 
of processes. In particular, they may be responsible for a large 
contribution to the quantity $\epsilon'/\epsilon$ measuring direct 
CP violation in $K\to\pi\pi$ decays \cite{epspr,eps1}. The 
importance of $B^\pm\to\pi K$ decays in the search for New Physics 
has been emphasized in \cite{me,Mati}, and some specific scenarios 
containing new isospin-violating contributions have been explored 
in \cite{CDK,anom}. 

In Section~\ref{sec:ops}, we discuss the effective weak 
Hamiltonian relevant to hadronic $B$ decays. In the presence of 
generic New Physics contributions, the basis of penguin operators 
has to be extended from the standard one in several aspects. We 
discuss the structure of the new operators and their scaling 
properties under a renormalization-group transformation from a 
high scale down to low energies. The theory of the rare hadronic 
decays $B^\pm\to\pi K$ is discussed in Section~\ref{sec:BpiK}, 
where we indicate how various observables in these decays can be 
used to test for physics beyond the SM. We derive model-independent
bounds on a ratio $R_*$ of CP-averaged $B^\pm\to\pi K$ branching 
ratios in the presence of New Physics, and show that the value
$\gamma_{\pi K}$ of the weak phase extracted in $B\to\pi K$ 
decays is extremely sensitive to isospin-violating New Physics
contributions. Because the theoretical analysis has only small 
hadronic uncertainties, potential New Physics effects 
can be detected even if they are 50\% smaller than the 
isospin-violating contributions present in the SM. In 
Sections~\ref{sec:tree} and \ref{sec:loop}, we calculate the 
Wilson coefficients of the $\bar b\to\bar s q\bar q$ penguin 
operators in a large class of extensions of the SM, including 
models with tree-level FCNC couplings of the $Z$ boson, extended 
gauge models, multi-Higgs models, and SUSY models with and without 
R-parity conservation. In each case, we explore which region of 
parameter space can be probed by measuring certain $B^\pm\to\pi K$ 
observables, and how big a departure from the SM predictions one 
can expect under realistic circumstances. Section~\ref{sec:concl} 
contains a summary of our results and the conclusions.

\section{Effective Hamiltonian for hadronic FCNC processes}
\label{sec:ops}

The effective weak Hamiltonians relevant to rare hadronic $B$ 
decays based on the quark transitions $\bar b\to\bar s q\bar q$ or 
$\bar b\to\bar d q\bar q$ have been discussed extensively in the
literature. Here we consider only the first case, adopting 
the notations of \cite{Heff}. A similar discussion (with obvious 
replacements of indices) would apply to the other case. In the SM, 
the result can be written in the compact form
\begin{equation}\label{Heff}
   {\cal H}_{\rm eff} = \frac{G_F}{\sqrt2}\,\sum_{q=u,c}
   \lambda_q\,\bigg[ \sum_{i=1,2} C_i(\mu)\,Q_i^q(\mu)
   + \sum_{i=3\dots 10} C_i(\mu)\,Q_i(\mu)
   + C_{8g}(\mu)\,Q_{8g}(\mu) \bigg] \,,
\end{equation}
where $\lambda_q=V_{qb}^* V_{qs}$ are combinations of CKM matrix 
elements obeying the unitarity relation 
$\lambda_u+\lambda_c+\lambda_t=0$, and $Q_i$ are local operators 
containing quark and gluon fields. Specifically, 
\begin{equation}\label{curcur}
   Q_1^q = (\bar b_\alpha q_\beta)_{V-A}\,
    (\bar q_\beta s_\alpha)_{V-A} \,, \qquad
   Q_2^q = (\bar b_\alpha q_\alpha)_{V-A}\,
    (\bar q_\beta s_\beta)_{V-A} \,,
\end{equation}
summed over color indices $\alpha$ and $\beta$, are the usual 
current--current operators induced by $W$-boson exchange, 
\begin{eqnarray}\label{QCDp}
   Q_3 &=& (\bar b_\alpha s_\alpha)_{V-A}\,\sum_q\,
   (\bar q_\beta q_\beta)_{V-A} \,, \nonumber\\
   Q_4 &=& (\bar b_\alpha s_\beta)_{V-A}\,\sum_q\,
    (\bar q_\beta q_\alpha)_{V-A} \,, \nonumber\\
   Q_5 &=& (\bar b_\alpha s_\alpha)_{V-A}\,\sum_q\,
    (\bar q_\beta q_\beta)_{V+A} \,, \nonumber\\
   Q_6 &=& (\bar b_\alpha s_\beta)_{V-A}\,\sum_q\,
    (\bar q_\beta q_\alpha)_{V+A} \,,
\end{eqnarray}
summed over the light flavors $q=u,d,s,c,b$, are referred to as 
QCD penguin operators, and
\begin{eqnarray}\label{EWp}
   Q_7 &=& \frac32\,(\bar b_\alpha s_\alpha)_{V-A}\,\sum_q\,e_q\,
    (\bar q_\beta q_\beta)_{V+A} \,, \nonumber\\
   Q_8 &=& \frac32\,(\bar b_\alpha s_\beta)_{V-A}\,\sum_q\,e_q\,
    (\bar q_\beta q_\alpha)_{V+A} \,, \nonumber\\
   Q_9 &=& \frac32\,(\bar b_\alpha s_\alpha)_{V-A}\,\sum_q\,e_q\,
    (\bar q_\beta q_\beta)_{V-A} \,, \nonumber\\
   Q_{10} &=& \frac32\,(\bar b_\alpha s_\beta)_{V-A}\,\sum_q\,e_q\,
    (\bar q_\beta q_\alpha)_{V-A} \,,
\end{eqnarray}
with $e_q$ denoting the electric charges of the quarks, are called 
electroweak penguin operators. The notation 
$(\bar q_1 q_2)_{V\pm A}$ implies $\bar q_1\gamma^\mu
(1\pm\gamma_5)q_2$. The terminology of QCD and electroweak 
penguins is slightly misleading insofar as the Wilson coefficients 
of the QCD penguin operators also receive small contributions from 
electroweak penguin and box diagrams. However, in the SM there are 
no strong-interaction contributions to the coefficients of the 
electroweak penguin operators. The operator $Q_{8g}
=\frac{g_s m_b}{8\pi^2}\,\bar b\,\sigma^{\mu\nu}(1-\gamma_5)
G_{\mu\nu}s$ in (\ref{Heff}) is the chromomagnetic dipole 
operator. The analogous electromagnetic dipole operator and 
semileptonic operators containing products of a quark current 
with a lepton current can be safely discarded from the effective 
Hamiltonian for hadronic $B$ decays.

In general, physics beyond the SM can induce a much larger set of 
penguin operators, and it is therefore unavoidable for our 
purposes to generalize the standard nomenclature reviewed above. 
However, for all the models we explore below it is sufficient to 
consider products of vector and/or axial vector currents only. We 
define a basis of such operators by
\begin{eqnarray}\label{basis}
   O_1^q &=& (\bar b_\alpha s_\alpha)_{V-A}\,
    (\bar q_\beta q_\beta)_{V+A} \,, \qquad
    O_2^q = (\bar b_\alpha s_\beta)_{V-A}\,
    (\bar q_\beta q_\alpha)_{V+A} \,, \nonumber\\
   O_3^q &=& (\bar b_\alpha s_\alpha)_{V-A}\,
    (\bar q_\beta q_\beta)_{V-A} \,, \qquad
    O_4^q = (\bar b_\alpha s_\beta)_{V-A}\,
    (\bar q_\beta q_\alpha)_{V-A} \,, \nonumber\\
   O_5^q &=& (\bar b_\alpha q_\alpha)_{V-A}\,
    (\bar q_\beta s_\beta)_{V+A} \,, \qquad
    O_6^q = (\bar b_\alpha q_\beta)_{V-A}\,
    (\bar q_\beta s_\alpha)_{V+A} \,,
\end{eqnarray}
and denote their Wilson coefficients by $c_i^q$. We implicitly 
assume a regularization scheme that preserves Fierz identities. 
In a general model, we also need operators of opposite chirality 
compared to the ones shown above, i.e., with $V-A\leftrightarrow 
V+A$ everywhere. We denote these operators by $\widetilde O_i^q$ 
and their coefficients by $\widetilde c_i^q$. Thus, the most 
general penguin Hamiltonian considered in this paper takes the 
form
\begin{equation}\label{Hpeng}
   {\cal H}_{\rm peng} = \frac{G_F}{\sqrt 2}\,\sum_i 
   \sum_q \left[ c_i^q(\mu)\,O_i^q(\mu)
   + \widetilde c_i^q(\mu)\,\widetilde O_i^q(\mu) \right] .
\end{equation}
To this, one has to add the current--current operators in 
(\ref{curcur}) and the chromo-magnetic dipole operator. It is 
implicitly understood that for the cases where $q=s$ or $b$ the 
operators $O_5^q$ and $O_6^q$ as well as $\widetilde O_5^q$ and 
$\widetilde O_6^q$ are omitted from the list of operators, because
they are Fierz-equivalent to the remaining ones.

The QCD and electroweak penguin operators present in the effective
weak Hamiltonian of the SM in(\ref{Heff}) are linear combinations
of the four operators $O_{1\dots 4}^q$. However, in a general 
model there may be additional penguin operators built out of
$O_5^q$, $O_6^q$, and of the opposite-chirality operators 
$\widetilde O_i^q$. Also, there may be operators which cannot be 
represented as linear combinations of QCD and electroweak 
penguin operators as shown in (\ref{QCDp}) and (\ref{EWp}), 
because in a general model the flavor symmetry 
among up- or down-type quarks can be violated. An example would 
be operators with flavor content $\bar sb(\bar d d-\bar s s)$, 
which are absent in the SM. Because such operators are of no 
relevance to our discussion they will not be explored any further 
here. For completeness, we show how the Wilson coefficients 
$C_{3\dots 10}$ of the SM penguin operators in (\ref{Heff}) can 
be expressed in terms of the coefficients $c_i^q$. Defining the 
linear combinations
\begin{equation}\label{cdefs}
   c_i^{\rm QCD} \equiv \frac{c_i^u+2c_i^d}{3} \,, \qquad
   c_i^{\rm EW} \equiv c_i^u-c_i^d \,,
\end{equation}
we have
\begin{eqnarray}
   -\lambda_t\,C_{3,4} &=& c_{3,4}^{\rm QCD} \,, \qquad
   -\lambda_t\,C_{5,6} = c_{1,2}^{\rm QCD} \,,
    \nonumber\\
   -\frac32\,\lambda_t\,C_i &=& c_{i-6}^{\rm EW} \,;
    \quad i=7,\dots,10 \,.
\end{eqnarray}

The isospin-violating effects induced by the coefficients 
$c_i^{\rm EW}$ and $\widetilde c_i^{\rm EW}$ are the main focus 
of this paper. In the SM, the matching conditions for the 
corresponding electroweak penguin coefficients at the weak scale, 
and to leading order in perturbation theory, are 
$C_8(m_W)=C_{10}(m_W)=0$, $C_7(m_W)\approx 0$, and \cite{Bur93}
\begin{equation}\label{C9match}
   C_9(m_W) \approx -\frac{\alpha}{12\pi}\,
   \frac{x_t}{\sin^2\!\theta_W} \left( 1 + \frac{3\ln x_t}{x_t-1}
   \right) ,
\end{equation}
where $x_t=(m_t/m_W)^2$. For simplicity, we show only the large 
electroweak contribution to $C_9(m_W)$ and omit a common, 
renormalization-scheme dependent electromagnetic contribution to 
$C_7(m_W)$ and $C_9(m_W)$, which is negligible compared with the 
contribution in (\ref{C9match}). In this approximation, the 
SM matching conditions for the electroweak penguin coefficients 
in our basis read 
\begin{eqnarray}\label{SMinit}
   c_3^{\rm EW,SM}(m_W) &=& \frac{\alpha}{8\pi}\,
    \frac{\lambda_t\,x_t}{\sin^2\!\theta_W} \left( 1
    + \frac{3\ln x_t}{x_t-1} \right)
    \approx -5.5\times 10^{-4} \,, \nonumber\\
   c_{i\ne 3}^{\rm EW,SM}(m_W) 
   &=& \widetilde c_i^{\rm EW,SM}(m_W) = 0 \,.
\end{eqnarray}
For the numerical estimate we have used $m_t=\overline{m}_t(m_t)
=170$\,GeV, $\alpha=1/129$, and $\lambda_t=V_{tb}^* V_{ts}=-0.04$. 
 
In Sections~\ref{sec:tree} and \ref{sec:loop}, we will calculate 
the Wilson coefficients $c_i^q$ at a high scale, which for 
simplicity will be identified with the electroweak scale. In 
phenomenological applications, however, one usually prefers 
working with coefficients renormalized at a low scale of order 
$m_b$. The two sets of coefficients are connected by a 
renormalization-group transformation \cite{Heff}. Here we discuss 
the QCD evolution of the electroweak penguin operators at the 
leading logarithmic order, neglecting next-to-leading corrections 
as well as QED corrections to the anomalous dimensions of the 
operators. Under QCD evolution, flavor-nonsinglet combinations of 
penguin operators mix into flavor-singlet combinations through 
diagrams in which two light quarks annihilate into a gluon, which 
then fragments into a pair of light quarks. However, there is no 
mixing of flavor-singlet operators into flavor-nonsinglet
ones. If we restrict ourselves to flavor-nonsinglet combinations 
of the operators $O_i^q$, each pair in the three lines in 
(\ref{basis}) obeys a separate matrix evolution equation. It 
follows that the coefficients $c_i^{\rm EW}$ mix pairwise under 
renormalization. For each pair of coefficients associated with 
$(V\mp A)\otimes(V\pm A)$ operators we obtain
\begin{eqnarray}\label{RGE-R}
   c_1^{\rm EW}(\mu) &=& \kappa^{-3/23}\,c_1^{\rm EW}(m_W) \,,
    \nonumber\\
   c_2^{\rm EW}(\mu) &=& \frac{\kappa^{24/23}-\kappa^{-3/23}}{3}\,
    c_1^{\rm EW}(m_W) + \kappa^{24/23}\,c_2^{\rm EW}(m_W) \,,
\end{eqnarray}
and similarly for $(c_5^{\rm EW},c_6^{\rm EW})$, where 
$\kappa=\alpha_s(\mu)/\alpha_s(m_W)$. For a pair of coefficients 
associated with $(V\mp A)\otimes(V\mp A)$ operators we obtain 
instead
\begin{eqnarray}\label{RGE-L}
   c_3^{\rm EW}(\mu) &=& \phantom{-}
    \frac{\kappa^{12/23}+\kappa^{-6/23}}{2}\,c_3^{\rm EW}(m_W)
    - \frac{\kappa^{12/23}-\kappa^{-6/23}}{2}\,
    c_4^{\rm EW}(m_W) \,, \nonumber\\
   c_4^{\rm EW}(\mu) &=& -\frac{\kappa^{12/23}-\kappa^{-6/23}}{2}\,
    c_3^{\rm EW}(m_W) + \frac{\kappa^{12/23}+\kappa^{-6/23}}{2}\,
    c_4^{\rm EW}(m_W) \,.
\end{eqnarray}
The coefficients $\widetilde c_i^q$ scale in the same way as 
the $c_i^q$. Since our main focus is on electroweak penguins and 
their generalizations beyond the SM, we will not discuss the more 
complicated evolution equations for the coefficients $c_i^q$ 
themselves, which can however readily be deduced from the 
literature \cite{Heff}.

\section{\boldmath 
Searching for New Physics with $B^\pm\to\pi K$ decays
\unboldmath}
\label{sec:BpiK}

In close correspondence with the different types of operators
in the effective weak Hamiltonian, one distinguishes three classes 
of flavor topologies relevant to $B\to\pi K$ decays, referred to 
as trees, QCD penguins and electroweak penguins. In the SM, the 
weak couplings associated with these topologies are known. From 
the measured branching ratios for the various $B\to\pi K$ decay 
modes it follows that the QCD penguins dominate the decay 
amplitudes \cite{Digh}, whereas trees and electroweak penguins 
are subleading and of a similar strength \cite{oldDesh}.
The theoretical description of the two charged modes 
$B^\pm\to\pi^\pm K^0$ and $B^\pm\to\pi^0 K^\pm$ exploits the fact 
that the amplitudes for these processes differ in a pure isospin 
amplitude $A_{3/2}$, defined as the matrix element of the 
isovector part of the effective Hamiltonian between a $B$ meson 
and the $\pi K$ isospin eigenstate with $I=\frac 32$. In the SM 
the parameters of this amplitude are determined, up to an overall 
strong-interaction phase $\phi$, in the limit of SU(3) flavor 
symmetry \cite{NR}. SU(3)-breaking corrections can be calculated 
in the factorization approximation \cite{Stech}, so that 
theoretical uncertainties enter only at the level of 
nonfactorizable SU(3)-breaking corrections to a subleading decay 
amplitude. Moreover, it has recently been shown that even these 
nonfactorizable corrections can be calculated in a model-independent
way up to terms that are power suppressed in $\Lambda/m_b$ and 
vanish in the heavy-quark limit \cite{fact}. 

\subsection{General parametrization of the decay amplitudes}

In the presence of New Physics, the analysis of $B^\pm\to\pi K$ 
decays becomes more complicated. A convenient and completely 
general parametrization of the two decay amplitudes is
\begin{eqnarray}\label{ampls}
   {\cal A}(B^+\to\pi^+ K^0) &=& P\,(1-i\rho\,e^{i\phi_\rho})
    \,,\nonumber\\
   -\sqrt2\,{\cal A}(B^+\to\pi^0 K^+) &=& P \Big[ 
    1-i\rho\,e^{i\phi_\rho} - \varepsilon_{3/2}\,e^{i\phi}
    (e^{i\gamma} - a\,e^{i\phi_a} - ib\,e^{i\phi_b}) \Big] \,,
\end{eqnarray}
where $P$ is the dominant penguin amplitude defined as the 
sum of all CP-conserving terms in the $B^\pm\to\pi^\pm K^0$ decay 
amplitudes, $\varepsilon_{3/2}$, $\rho$, $a$, $b$ are real 
hadronic parameters, and $\phi$, $\phi_\rho$, $\phi_a$, $\phi_b$ 
are strong-interaction phases. The weak phase 
$\gamma=\mbox{arg}(V_{ub}^*)$ and the terms $i\rho$ and $ib$ 
change sign under a CP transformation, whereas all other 
parameters stay invariant. The terms proportional to 
$\varepsilon_{3/2}$ in (\ref{ampls}) parameterize the isospin 
amplitude $A_{3/2}$. The contribution proportional to $e^{i\gamma}$ 
comes from the matrix elements of the current--current operators 
$Q_1^u$ and $Q_2^u$ in the effective Hamiltonian, which mediate the 
tree process $\bar b\to\bar u u\bar s$. The quantities $a$ and $b$ 
parameterize the effects of electroweak penguins. It is crucial that 
only isospin-violating terms can contribute to the amplitude 
$A_{3/2}$. All isospin-conserving contributions reside in $P$ and 
$\rho$. 

Let us discuss the various terms entering the decay amplitudes in
detail. The parameter $\varepsilon_{3/2}$ characterizes the 
relative strength of tree and QCD penguin contributions. 
Information about it can be derived by using SU(3) flavor symmetry 
to relate the tree contribution to the isospin amplitude $A_{3/2}$ 
to the corresponding contribution in the decay $B^+\to\pi^+\pi^0$. 
Since the final state $\pi^+\pi^0$ has isospin $I=2$ (because of 
Bose symmetry), the amplitude for the latter process does not 
receive any contribution from QCD penguins. Moreover, in the SM 
electroweak penguins in $\bar b\to\bar d q\bar q$ transitions are 
negligible, and thus only the tree topology contributes to the 
$B^+\to\pi^+\pi^0$ decay amplitude. In our analysis we make the 
plausible assumption that potential New Physics contributions to 
this amplitude can be neglected.\footnote{If this were not the 
case, the New Physics impact on $\bar\varepsilon_{3/2}$ would 
provide us with another handle on non-standard isospin-violating 
effects.} 
Even if new electroweak penguin effects would be of comparable
strength in $\bar b\to\bar s q\bar q$ and $\bar b\to\bar d q\bar q$
transitions, the latter would have to compete with a 
Cabibbo-enhanced tree amplitude in order to be significant 
in $B^+\to\pi^+\pi^0$ decays. We then find that \cite{NR,me}
\begin{equation}\label{eps}
   \bar\varepsilon_{3/2} 
   \equiv \frac{\varepsilon_{3/2}}{\sqrt{1+\rho^2}}
   = \sqrt2\,R_{\rm SU(3)} \left|\frac{V_{us}}{V_{ud}}\right|
   \left[
   \frac{\mbox{B}(B^+\to\pi^+\pi^0)+\mbox{B}(B^-\to\pi^-\pi^0)}
        {\mbox{B}(B^+\to\pi^+ K^0)+\mbox{B}(B^-\to\pi^-\bar K^0)}
   \right]^{1/2} .
\end{equation}
SU(3)-breaking corrections are described by the factor 
$R_{\rm SU(3)}=1.22\pm 0.05$, which can be calculated in a 
model-independent way using the QCD factorization theorem of 
\cite{fact}. The quoted error is an estimate of the theoretical 
uncertainty due to uncontrollable corrections of
$O(\frac{1}{N_c}\frac{m_s}{m_b})$. Using preliminary data
reported by the CLEO Collaboration \cite{CLEO} to evaluate the
ratio of branching ratios in (\ref{eps}), we obtain
\begin{equation}\label{epsval}
   \bar\varepsilon_{3/2} = 0.21\pm 0.06_{\rm exp}
   \pm 0.01_{\rm th} \,.
\end{equation}
With a better measurement of the branching ratios the uncertainty 
in $\bar\varepsilon_{3/2}$ will be reduced significantly.

The parameter $\rho$ in (\ref{ampls}) parameterizes the sum of all 
CP-violating contributions to the $B^+\to\pi^+ K^0$ decay 
amplitude. In the presence of New Physics, those contributions 
could arise from QCD as well as electroweak penguin operators. 
Note that the CP-conserving part of such terms is absorbed, by 
definition, into the quantity $P$. We will not attempt a 
theoretical calculation of this quantity (which is difficult even 
in the SM) and only consider observables that are independent of 
$P$. In the SM, $\rho\simeq\varepsilon_a\sin\gamma$ \cite{me} 
describes a small contribution induced by final-state 
rescattering from tree or annihilation diagrams 
\cite{Blok97,BFM98,Ge97,Ne97,Fa97,At97}. In the heavy-quark limit,
the parameter $\varepsilon_a$ can be calculated and is found to 
be of order $-2\%$ \cite{fact}. 

Finally, in the SM the parameter $b$ vanishes, while 
\begin{equation}\label{delval}
   a\,e^{i\phi_a} = \delta_{\rm EW} 
   = (0.64\pm 0.09)\times\frac{0.085}{|V_{ub}/V_{cb}|}
\end{equation}
is calculable in terms of fundamental parameters \cite{NR,me,Fl96}. 
Up to some small SU(3)-breaking corrections, $\delta_{\rm EW}$ is 
given by the Wilson coefficient $c_3^{\rm EW}(m_W)$ in 
(\ref{SMinit}) divided by $-|\lambda_u|$. There are no additional 
hadronic uncertainties in this estimate in the SM. In 
particular, the strong-interaction phase $\phi_a$ is bounded to 
be less than a few degrees and can be neglected for all practical
purposes \cite{me}. In a general model, the parameters $a$ and $b$ 
depend on the values of the penguin coefficients $c_i^{\rm EW}$ 
and $\widetilde c_i^{\rm EW}$ as well as on the hadronic matrix 
elements of the corresponding operators evaluated between a $B$ 
meson and the $\pi K$ isospin state with $I=\frac32$. Since our 
intention here is to look for New Physics effects rather than 
doing precision calculations, it will be sufficient for our 
purposes to evaluate these matrix elements in a given New Physics 
scenario using the naive factorization approximation and 
neglecting small SU(3)-breaking effects.\footnote{According to 
\protect\cite{fact}, naive factorization gives the leading term 
in the heavy-quark limit.} 
Then the strong-interaction phases $\phi_a$ and $\phi_b$ vanish,
and we obtain 
\begin{eqnarray}\label{abres}
   |\lambda_u|\,(a+ib) &\approx&
    - \left( \bar c_3^{\rm EW} + \bar c_4^{\rm EW} \right) 
    + \frac23\,\kappa^{3/23}
    \left( \bar c_1^{\rm EW} + \bar c_5^{\rm EW} \right)
    \nonumber\\
   &&\mbox{}- \frac{3\chi-1}{4}\,\kappa^{30/23}
    \bigg( \bar c_2^{\rm EW} + \frac13\,\bar c_1^{\rm EW} 
    + \bar c_6^{\rm EW} + \frac13\,\bar c_5^{\rm EW} \bigg) \,,
\end{eqnarray}
where $\bar c_i^{\rm EW}\equiv c_i^{\rm EW}(m_W)
-\widetilde c_i^{\rm EW}(m_W)$, and
\begin{equation}
   \chi = \frac{2 m_K^2}{(m_s+m_d)\,m_b}
   = \frac{2 m_\pi^2}{(m_u+m_d)\,m_b} \,.
\end{equation}
Parity invariance implies that the relevant hadronic matrix 
elements of the operators $\widetilde O_i$ have the opposite
sign compared with those of the operators $O_i$. The fact that 
$\bar c_3^{\rm EW}$ and $\bar c_4^{\rm EW}$ enter with the 
coefficient $-1$ in (\ref{abres}) is a model-independent result 
free of hadronic uncertainties, irrespective of whether these 
coefficients receive New Physics contributions or not. It follows 
because the isovector components of the penguin operators 
$O_3^q$ and $O_4^q$ can be related to the usual current--current 
operators by a Fierz transformation \cite{Ne97,Fl96}. For the 
numerical analysis we choose the renormalization scale $\mu=m_b$ 
and take $\kappa=\alpha_s(m_b)/\alpha_s(m_W)=1.83$ and $\chi=1.18$, 
yielding
\begin{equation} \label{abformula}
   |\lambda_u|\,(a+ib) \approx
   - (\bar c_3^{\rm EW} + \bar c_4^{\rm EW})
   + 0.26 (\bar c_1^{\rm EW} + \bar c_5^{\rm EW})
   - 1.40 (\bar c_2^{\rm EW} + \bar c_6^{\rm EW}) \,.
\end{equation}

Next we consider the New Physics contributions to the parameter 
$\rho$. In the factorization approximation, we find 
\begin{equation}\label{rhobound}
   \frac{\rho}{\sqrt{1+\rho^2}}
   \approx \frac{3\bar\varepsilon_{3/2}}{4|\lambda_u|}\,
   \mbox{Im}\Big[ 
   (\bar c_5^d - \bar c_4^d - \chi\,\bar c_2^d)
   + \frac13\,(\bar c_6^d - \bar c_3^d - \chi\,\bar c_1^d)
   - a_{8g}\,C_{8g} \Big] \,,
\end{equation}
where $\bar c_i^d\equiv c_i^d(m_b)-\widetilde c_i^d(m_b)$. 
Because the QCD evolution of the Wilson coefficients $c_i^d$ and 
$\widetilde c_i^d$ is complicated, we prefer to present the 
result in terms of coefficients renormalized at 
the scale $m_b$. Note that by pulling out a factor of 
$\bar\varepsilon_{3/2}$ on the right-hand side we avoid the 
difficulty of calculating the overall penguin amplitude $P$ in 
(\ref{ampls}). The contribution of the chromomagnetic dipole 
operator is formally of next-to-leading order in $\alpha_s$ and 
thus could be dropped; however, we keep it because in some New 
Physics models the coefficient $C_{8g}$ can be enhanced with 
respect to its SM value by an order of magnitude 
\cite{glue1,George,Alex,CGGi}. Using the QCD factorization 
approach of \cite{fact}, we find that
\begin{equation}
   a_{8g} = \frac{2\alpha_s}{3\pi}\,(1+\chi)(-\lambda_t)
   \approx 4\times 10^{-3} \,.
\end{equation}
Note that the magnitude of the left-hand side in (\ref{rhobound}) 
is bounded by unity. This implies a nontrivial upper bound on the 
possible CP-violating New Physics contributions to the penguin 
operators, given by
\begin{equation}\label{rhonumeric}
   \left|\,\mbox{Im}\Big[ 
   (\bar c_5^d - \bar c_4^d - \chi\,\bar c_2^d)
   + \frac13\,(\bar c_6^d - \bar c_3^d - \chi\,\bar c_1^d)
   - a_{8g}\,C_{8g} \Big] \right| < 
   \frac{4|\lambda_u|}{3\bar\varepsilon_{3/2}} \,.
\end{equation}
Using $|V_{ub}/V_{cb}|=0.085\pm 0.015$, corresponding to 
$|\lambda_u|=(7.5\pm 1.3)\times 10^{-4}$, and taking for 
$\bar\varepsilon_{3/2}$ the value in (\ref{epsval}), we find that
the right-hand side of this bound is less than $7.4\times 10^{-3}$
at 90\% confidence level. For comparison, we note that in the SM the 
magnitude of the combination of penguin coefficients entering above 
is about $3\times 10^{-3}$. However, due to the smallness of the 
weak phase of $\lambda_t$ the imaginary part of this combination is 
much smaller.

\subsection{Model-independent bounds on $R_*$ in the presence of 
New Physics}

The most important observable in the exploration of New Physics 
effects in $B\to\pi K$ decays is the ratio of the CP-averaged 
branching ratios for the decays $B^\pm\to\pi^\pm K^0$ and 
$B^\pm\to\pi^0 K^\pm$, given by
\begin{equation}\label{Rstexp}
   R_* =
   \frac{\mbox{B}(B^+\to\pi^+ K^0)+\mbox{B}(B^-\to\pi^-\bar K^0)}
        {2[\mbox{B}(B^+\to\pi^0 K^+)+\mbox{B}(B^-\to\pi^0 K^-)]}
   = 0.75\pm 0.28 \,,
\end{equation}
where the quoted experimental value is derived from data reported
by the CLEO Collaboration \cite{CLEO}. It will often be convenient
to consider a related quantity defined as
\begin{equation}\label{XRval}
   X_R = \frac{\sqrt{R_*^{-1}}-1}{\bar\varepsilon_{3/2}} 
   = 0.72\pm 0.98_{\rm exp}\pm 0.03_{\rm th} \,.
\end{equation}
Because of the theoretical factor $R_{\rm SU(3)}$ entering the 
definition of $\bar\varepsilon_{3/2}$ in (\ref{eps}) this is, 
strictly speaking, not an observable. However, the irreducible 
theoretical uncertainty in $X_R$ is so much smaller than the 
present experimental error that it is justified to treat this
quantity as an observable. The advantage of presenting our 
results in terms of $X_R$ rather than $R_*$ is that the leading 
dependence on $\bar\varepsilon_{3/2}$ cancels out (see below). 
Also, some experimental errors cancel in the ratio in (\ref{XRval}) 
\cite{Frank}.

When writing theoretical expressions for the quantities $R_*$ and
$X_R$ we eliminate the two parameters $\varepsilon_{3/2}$ and 
$\rho$ in favor of the measurable parameter $\bar\varepsilon_{3/2}$ 
and a ``weak phase'' $\varphi\in[-90^\circ,90^\circ]$ defined by
\begin{equation}
   \sin\varphi = \frac{\rho}{\sqrt{1+\rho^2}} \,, \qquad
   \cos\varphi = \frac{1}{\sqrt{1+\rho^2}} \,.
\end{equation}
The most direct way of probing this phase is via the direct CP 
asymmetry in the decays $B^\pm\to\pi^\pm K^0$, which is given by
\begin{equation}
   A_{\rm CP}(\pi^+ K^0) =
   \frac{\mbox{B}(B^+\to\pi^+ K^0)-\mbox{B}(B^-\to\pi^-\bar K^0)}
        {\mbox{B}(B^+\to\pi^+ K^0)+\mbox{B}(B^-\to\pi^-\bar K^0)} 
   = \sin2\varphi\sin\phi_\rho \,.
\end{equation}
In the SM, $\sin2\varphi\approx 2\varepsilon_a\sin\gamma$ is of 
order a few percent, and with realistic values for $\phi_\rho$
of order $10^\circ$--$20^\circ$ \cite{fact} one expects a very 
small CP asymmetry. However, in New Physics 
scenarios with new CP-violating couplings $\sin 2\varphi$ may be 
significantly larger than in the SM. An experimental finding that 
$A_{\rm CP}(\pi^+ K^0)=O(10\%)$ would constitute strong evidence 
for the existence of such an effect.

The exact theoretical expression for $R_*$ is 
\begin{eqnarray}\label{Rstar}
   R_*^{-1} &=& 1 + 2\bar\varepsilon_{3/2}\cos\varphi
    \Big[ a\cos(\phi+\phi_a) - \cos\gamma\cos\phi \Big] \nonumber\\
   &&\mbox{}- 2\bar\varepsilon_{3/2}\sin\varphi
    \Big[ b\cos(\phi+\phi_b-\phi_\rho)
    - \sin\gamma\cos(\phi-\phi_\rho) \Big] \nonumber\\
   &&\mbox{}+ \bar\varepsilon_{3/2}^2 \left( 1 
    - 2a\cos\gamma\cos\phi_a - 2b\sin\gamma\cos\phi_b 
    + a^2 + b^2 \right) .
\end{eqnarray}
In the SM $b=0$, $a=\delta_{\rm EW}$, $\phi_a=0$, and
$\varphi\approx 0$ to very good approximation. Therefore
\begin{eqnarray}\label{Rstbound}
   R_*^{-1} &=& 1 + 2\bar\varepsilon_{3/2}\,
    (\delta_{\rm EW}-\cos\gamma)\cos\phi 
    + \bar\varepsilon_{3/2}^2
    (1 - 2\delta_{\rm EW}\cos\gamma + \delta_{\rm EW}^2)
    \nonumber\\
   &\le& \left( 1 + \bar\varepsilon_{3/2}\,
    |\delta_{\rm EW}-\cos\gamma| \right)^2 
    + \bar\varepsilon_{3/2}^2\sin^2\!\gamma \,.
\end{eqnarray}
In the second step we have used the fact that $|\cos\phi|\le 1$
to obtain an upper bound on $R_*^{-1}$. Similarly, a lower bound
can be derived, which is obtained by changing the sign of 
$\bar\varepsilon_{3/2}$ in the above inequality. These bounds 
imply nontrivial constraints on $\cos\gamma$ provided that $R_*$ 
differs from 1 by a significant amount. In 
Figure~\ref{fig:SMbound}, we show the resulting lower and upper 
bounds on the quantity $X_R$ versus $\gamma$, obtained by scanning 
the input parameters in the intervals $0.15<\bar\varepsilon_{3/2}
<0.27$ and $0.49<\delta_{\rm EW}<0.79$. The latter value is 
obtained by using $|V_{ub}/V_{cb}|=0.085\pm 0.015$ in 
(\ref{delval}). The dependence on the variation of 
$\bar\varepsilon_{3/2}$ is so small that it would almost be 
invisible on the scale of the plot. Note that the extremal values 
$R_*$ can take in the SM are such that $|X_R|\le(1+\delta_{\rm EW})$ 
irrespective of the value of $\gamma$. A value exceeding this limit 
would be a clear signal for New Physics \cite{me,Mati}. In view of 
the present large error on $X_R$, this is still a realistic 
possibility. Because the upper and lower bounds on $X_R$ are, to

\EPSFIGURE{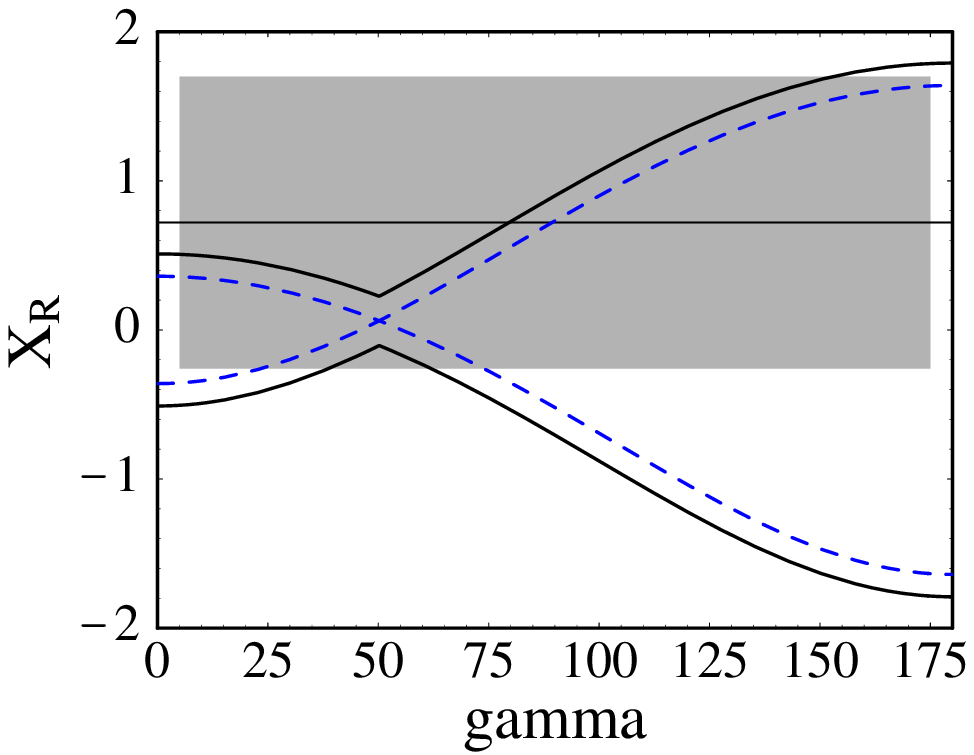,width=7.2cm} 
{\label{fig:SMbound}
Upper and lower bounds on $X_R$ versus $|\gamma|$ (in degrees) 
in the SM. The dashed line shows the bounds obtained using the 
central values of the parameters $\bar\varepsilon_{3/2}$ and 
$\delta_{\rm EW}$. The band shows the current experimental value
of $X_R$ with its $1\sigma$ variation.}

\noindent
a very good approximation, symmetric 
around $X_R=0$, we will from now on only show the upper bounds,
since $X_R>0$ is the region favored by experiment. Also, since the 
bounds change little under variation of $\delta_{\rm EW}$ and
$\bar\varepsilon_{3/2}$, we will work with the central values 
$\delta_{\rm EW}=0.64$ and $\bar\varepsilon_{3/2}=0.21$. By the 
time the experimental data will be sufficiently precise to perform 
the New Physics searches proposed in this work, the errors on 
these parameters are likely to be reduced by a significant amount.

Let us first discuss the case where New Physics induces 
arbitrary CP-vio\-la\-ting contributions to the $B\to\pi K$
decay amplitudes, while preserving isospin symmetry. Then the 
only change with respect to the SM would be that the weak phase 
$\varphi$ may no longer be negligible. We obtain
\begin{eqnarray}\label{Rstphi}
   R_*^{-1} &=& 1 + 2\bar\varepsilon_{3/2} \Big[
    \cos\varphi\,(\delta_{\rm EW}-\cos\gamma)\cos\phi
    + \sin\varphi\sin\gamma\cos(\phi-\phi_\rho) \Big] \nonumber\\
   &&\mbox{}+ \bar\varepsilon_{3/2}^2 
    (1 - 2\delta_{\rm EW}\cos\gamma + \delta_{\rm EW}^2) 
    \nonumber\\
   &\le& 1 + 2\bar\varepsilon_{3/2} \Big[ 
    \cos\varphi\,|\delta_{\rm EW}-\cos\gamma|
    + |\sin\varphi\sin\gamma| \Big] \nonumber\\
   &&\mbox{}+ \bar\varepsilon_{3/2}^2 
    (1 - 2\delta_{\rm EW}\cos\gamma + \delta_{\rm EW}^2) 
    \nonumber\\
   &\le& \left( 1 + \bar\varepsilon_{3/2}
   \sqrt{1 - 2\delta_{\rm EW}\cos\gamma + \delta_{\rm EW}^2}
   \right)^2 \,.
\end{eqnarray}
In deriving the upper bounds we have varied the strong-interaction 
phases $\phi$ and $\phi_\rho$ independently. Analogous lower 
bounds are obtained as previously by changing the sign of 
$\bar\varepsilon_{3/2}$. The last inequality in (\ref{Rstphi}) is 
remarkable in that it holds for arbitrary isospin-conserving New 
Physics effects no matter how large they are. The corresponding 
bound on $|X_R|$ is
\begin{equation}\label{phiarb}
   |X_R| \le 
   \sqrt{1 - 2\delta_{\rm EW}\cos\gamma + \delta_{\rm EW}^2}
   \le 1+\delta_{\rm EW} \,.
\end{equation}
Note that the extremal value is the same as in the SM, i.e., 
isospin-conserving New Physics effects cannot lead to a value of 
$|X_R|$ exceeding $1+\delta_{\rm EW}$. In the left-hand plot in 
Figure~\ref{fig:case12} we show the upper bound on $X_R$ versus 
$\gamma$ in the SM, and for New Physics scenarios with different 
values of $\varphi$. The three choices of $\varphi$ shown 
correspond to $|\rho|\approx 0.27$, 0.58 and 1. The gray curve
shows the upper bound obtained by varying $\varphi$. We observe 
that isospin-conserving New Physics can enhance the value of
$X_R$ relative to the SM, but only by a moderate amount. 

\FIGURE{
\epsfig{file=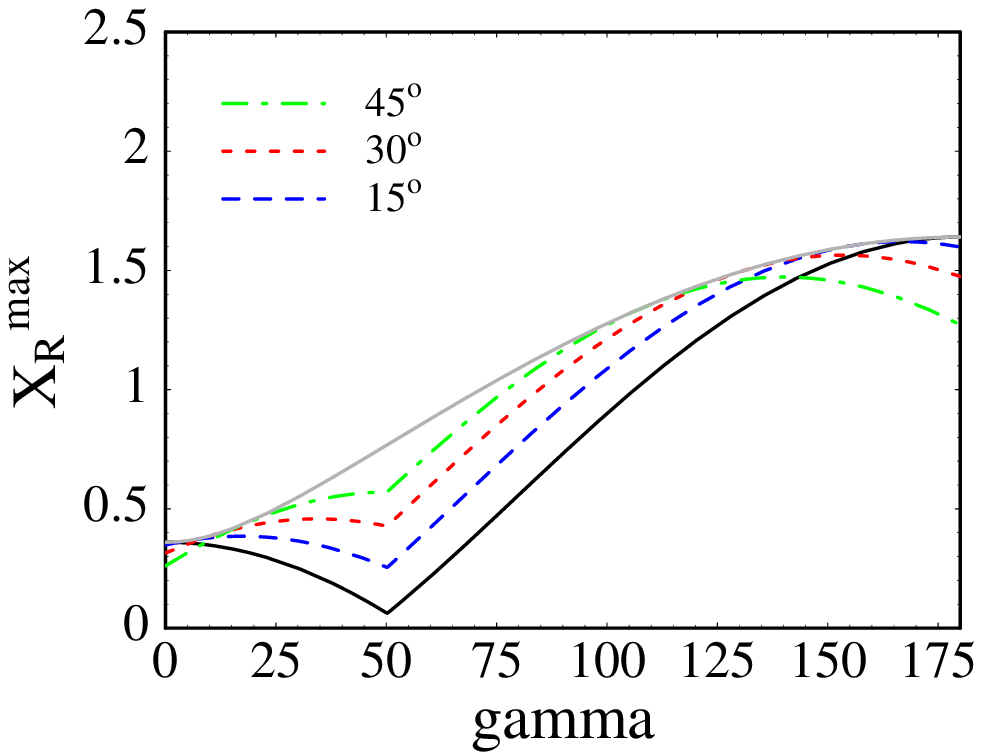,width=7cm}
\epsfig{file=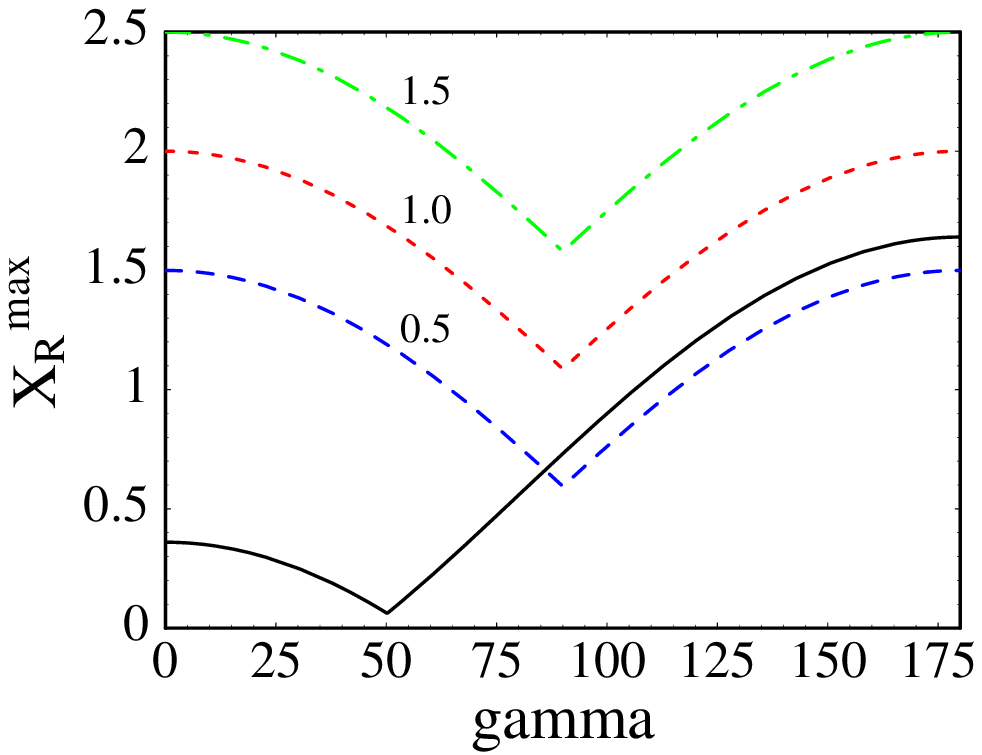,width=7cm} 
\caption{\label{fig:case12}
Upper bound on $X_R$ versus $|\gamma|$ for the SM (solid) and
with different New Physics contributions. Left: Isospin-conserving 
New Physics with $|\varphi|=15^\circ$ (dashed), $30^\circ$ 
(short-dashed), and $45^\circ$ (dashed-dotted). The gray curve 
shows the bound (\protect\ref{phiarb}) valid for arbitrary 
$\varphi$. Right: Isospin-violating but CP-conserving New Physics
with $|a|=0.5$ (dashed), 1 (short-dashed), and 1.5 
(dashed-dotted).}}

Next we consider New Physics effects that violate isospin symmetry, 
but we first restrict ourselves to the important subclass of 
models which do not contain significant new CP-violating phases. 
Then $\varphi$ and $b$ still vanish, and we obtain
\begin{eqnarray}\label{Rsta}
   R_*^{-1} &=& 1 + 2\bar\varepsilon_{3/2} \Big[
    a\cos(\phi+\phi_a) - \cos\gamma\cos\phi \Big] 
    + \bar\varepsilon_{3/2}^2 
    (1 - 2a\cos\phi_a\cos\gamma + a^2) \nonumber\\
   &\le& \left[ 1 + \bar\varepsilon_{3/2} \left(
    |a|+|\cos\gamma| \right) \right]^2
    + \bar\varepsilon_{3/2}^2\sin^2\!\gamma \,.
\end{eqnarray}
Again, a lower bound can be obtained by changing the sign of
$\bar\varepsilon_{3/2}$. In contrast to the previous case, now 
the maximal value of $|X_R|$ is given by $1+|a|$ and thus can
exceed the SM bound provided that $|a|>\delta_{\rm EW}$. This is 
illustrated in the right-hand plot in Figure~\ref{fig:case12}, 
where we show the resulting upper bound on $X_R$ versus $\gamma$ 
for different values of $|a|$. In contrast with the case of
isospin-conserving New Physics, even a moderate enhancement
of the coefficient $a$ corresponding to a 10\%--20\% change in 
the decay amplitudes can lead to a significant increase of the 
upper bound on $X_R$. 

If both isospin-violating and isospin-conserving New Physics
effects are present and involve new CP-violating phases, the 
analysis becomes more complicated. Still, it is possible to 
derive from (\ref{Rstar}) a series of bounds on $R_*^{-1}$. We 
find
\begin{eqnarray}
   R_*^{-1} &\le& 1 + 2\bar\varepsilon_{3/2} \Big[
    \cos\varphi\,(|a|+|\cos\gamma|)
    + |\sin\varphi|\,(|b|+|\sin\gamma|) \Big] \nonumber\\
   &&\mbox{}+ \bar\varepsilon_{3/2}^2 \Big[
    (|a|+|\cos\gamma|)^2 + (|b|+|\sin\gamma|)^2 \Big] \nonumber\\
   &\le& \left[ 1 + \bar\varepsilon_{3/2}
    \sqrt{(|a|+|\cos\gamma|)^2 + (|b|+|\sin\gamma|)^2} \right]^2 
    \nonumber\\
   &\le& \left[ 1 + \bar\varepsilon_{3/2} \left( 1
   + \sqrt{a^2 + b^2} \right) \right]^2 \,,
\end{eqnarray}
where in the second and third steps we have eliminated $\varphi$ 
and $\gamma$, respectively. As before a series of lower bounds is 
obtained by changing the sign of $\bar\varepsilon_{3/2}$. The 
corresponding bounds on $X_R$ are
\begin{equation}\label{abbound}
   |X_R| \le \sqrt{(|a|+|\cos\gamma|)^2 + (|b|+|\sin\gamma|)^2}
   \le 1 + \sqrt{a^2 + b^2}
   \le \frac{2}{\bar\varepsilon_{3/2}} + X_R \,,
\end{equation}
where the last inequality is relevant only in cases where
$\sqrt{a^2 + b^2}\gg\delta_{\rm EW}$. With the current values for
$\bar\varepsilon_{3/2}$ and $X_R$, the right-hand side is less 
than 15 at 90\% confidence level. The important point to note 
is that in the most general case, where $b$ and $\rho$ are
nonzero, the maximal value $X_R$ can take is no longer restricted 
to occur at the endpoints $\gamma=0^\circ$ or $180^\circ$, which 
are disfavored by the global analysis of the unitarity triangle 
\cite{Yuvi}. Rather, the maximal value $X_R^{\rm max}
=1+\sqrt{a^2 + b^2}$ now occurs at
\begin{equation}
   |\tan\gamma| = |\rho| = \bigg| \frac{b}{a} \bigg| \,.
\end{equation}

\EPSFIGURE{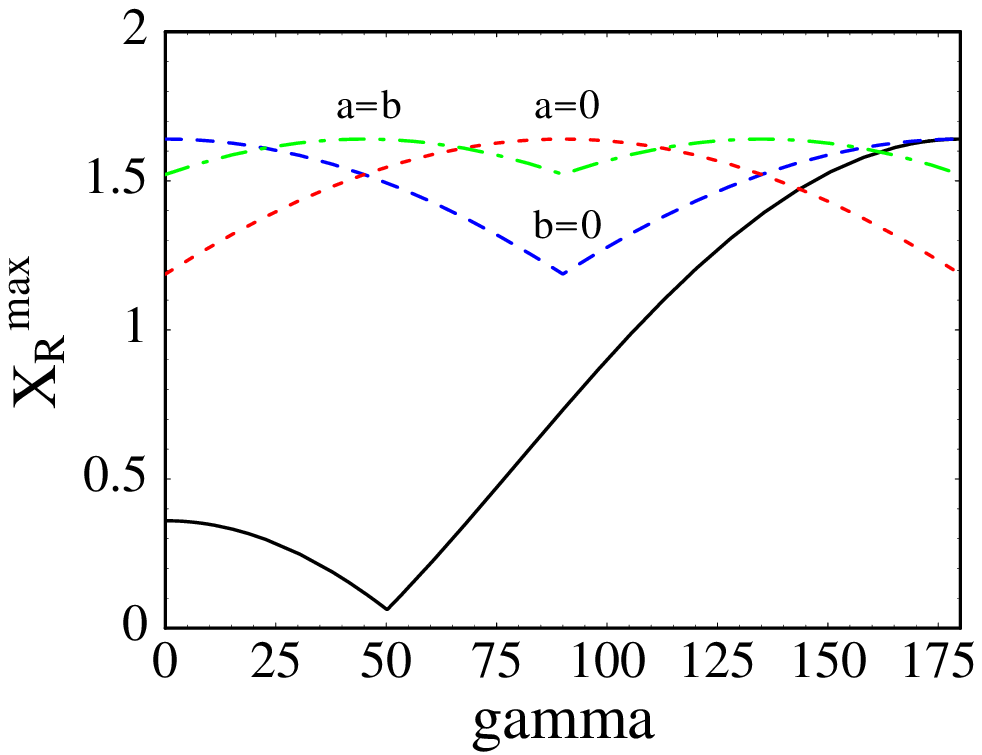,width=7.2cm} 
{\label{fig:case3}
Upper bound on $X_R$ versus $|\gamma|$ for arbitrary New Physics 
contributions satisfying $a^2+b^2=1$. The curves correspond to
$a=1$ and $b=0$ (dashed), $a=b=1/\sqrt{2}$ (dashed-dotted), 
and $a=0$ and $b=1$ (short-dashed). The SM bound is shown by 
the solid line.}

\noindent
The situation is illustrated in Figure~\ref{fig:case3}, where we 
show the upper bound on $X_R$ for arbitrary New Physics 
contributions satisfying $a^2+b^2=1$, but for different values of 
$a$ and $b$ chosen such that the maximum occurs at the endpoints 
$|\gamma|=0^\circ$ or $180^\circ$ ($b=0$), at the intermediate 
points $|\gamma|=45^\circ$ or $135^\circ$ ($a=b$), and in the 
center at $|\gamma|=90^\circ$ ($a=0$). The corresponding values of 
the New Physics parameter $\rho$ required to reach the maximum are 
$|\rho|=0$, 1, and $\infty$ (i.e., $|\rho|\gg 1$), respectively.

The present experimental value of $X_R$ in (\ref{XRval}) has too 
large an error to determine whether there is any deviation from 
the SM prediction. If $X_R$ turns out to be larger than 1 (i.e., 
only about one third of a standard deviation above its 
current central value), then an interpretation of this result in 
the SM would require a large value $|\gamma|>96^\circ$ (see 
Figure~\ref{fig:SMbound}), which may be difficult to accommodate. 
This may be taken as evidence for New Physics. If $X_R>1.3$, one 
could go a step further and conclude that this New Physics must 
necessarily violate isospin.

\subsection{New Physics effects on the determination of $\gamma$}

A value of the observable $R_*$ which violates the SM bound 
(\ref{Rstbound}) would be an exciting hint for new
isospin-violating penguin contributions from New Physics. With 
the current central value of $R_*$ derived from CLEO data this 
is still a realistic possibility. However, even if a more precise 
measurement will give a value that is consistent with the SM 
bound, $B^\pm\to\pi K$ decays still provide an excellent testing 
ground for physics beyond the SM. In the SM, the weak phase 
$\gamma$, along with the strong-interaction phase $\phi$, can be 
determined up to discrete ambiguities by combining measurements 
of $R_*$ and an asymmetry $\widetilde A$, which is defined as a 
linear combination of the 
direct CP asymmetries in the two $B^\pm\to\pi K$ decay channels 
\cite{NR,me}. The discrete ambiguities can be resolved using
information on the strong-interaction phase $\phi$ from theoretical
approaches such as the QCD factorization theorem derived in
\cite{fact}. The theoretical uncertainty on the value of $\gamma$
is typically of order $10^\circ$. Although New Physics may not be 
exotic enough to lead to a violation of the general bounds derived 
in the previous section, it may still cause a significant shift in 
the extracted value of $\gamma$. This may lead to inconsistencies 
when the value $\gamma_{\pi K}$ extracted in $B^\pm\to\pi K$ decays 
is compared with determinations of $\gamma$ using other information. 

A global fit of the unitarity 
triangle combining information from semi\-leptonic $B$ decays, 
$B$--$\bar B$ mixing, CP violation in the kaon system, and 
mixing-induced CP violation in $B\to J/\psi\,K_S$ decays will 
provide information on $\gamma$, which in a few years will 
determine its value within a rather narrow range \cite{BaBar,Yuvi}. 
Such an indirect determination could be comple\-men\-ted by direct 
measurements of $\gamma$ using, e.g., $B\to D K^{(*)}$ decays 
\cite{Soffer}, or using the triangle relation 
$\gamma=180^\circ-\alpha-\beta$ combined with a measurement of 
$\alpha$ in $B\to\pi\pi$ or $B\to\pi\rho$ decays \cite{BaBar}. In 
our discussion below we will assume that a discrepancy between the 
``true'' $\gamma=\mbox{arg}(V_{ub}^*)$ and the value 
$\gamma_{\pi K}$ extracted in $B^\pm\to\pi K$ decays of more than 
$25^\circ$ will be observable after a few years of operation at the 
$B$ factories. This will set the benchmark for sensitivity to 
New Physics effects.

\FIGURE{
\epsfig{file=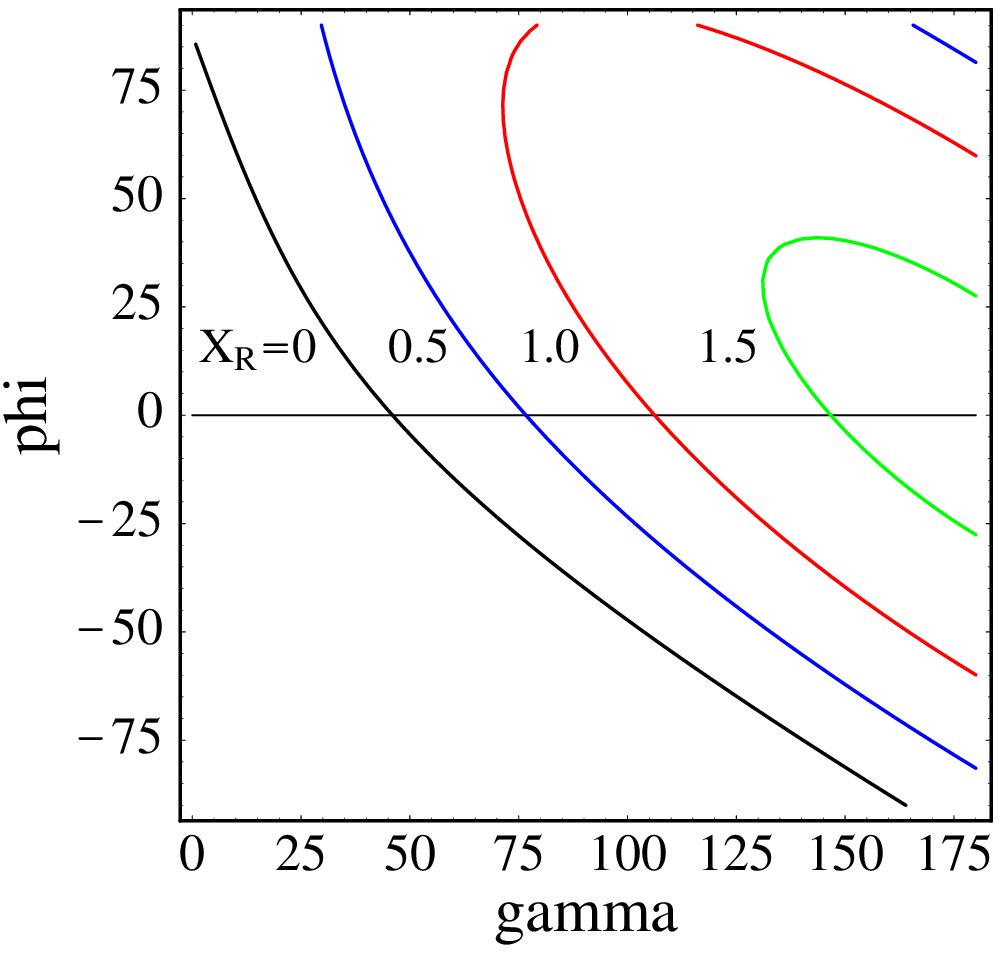,width=7cm}
\epsfig{file=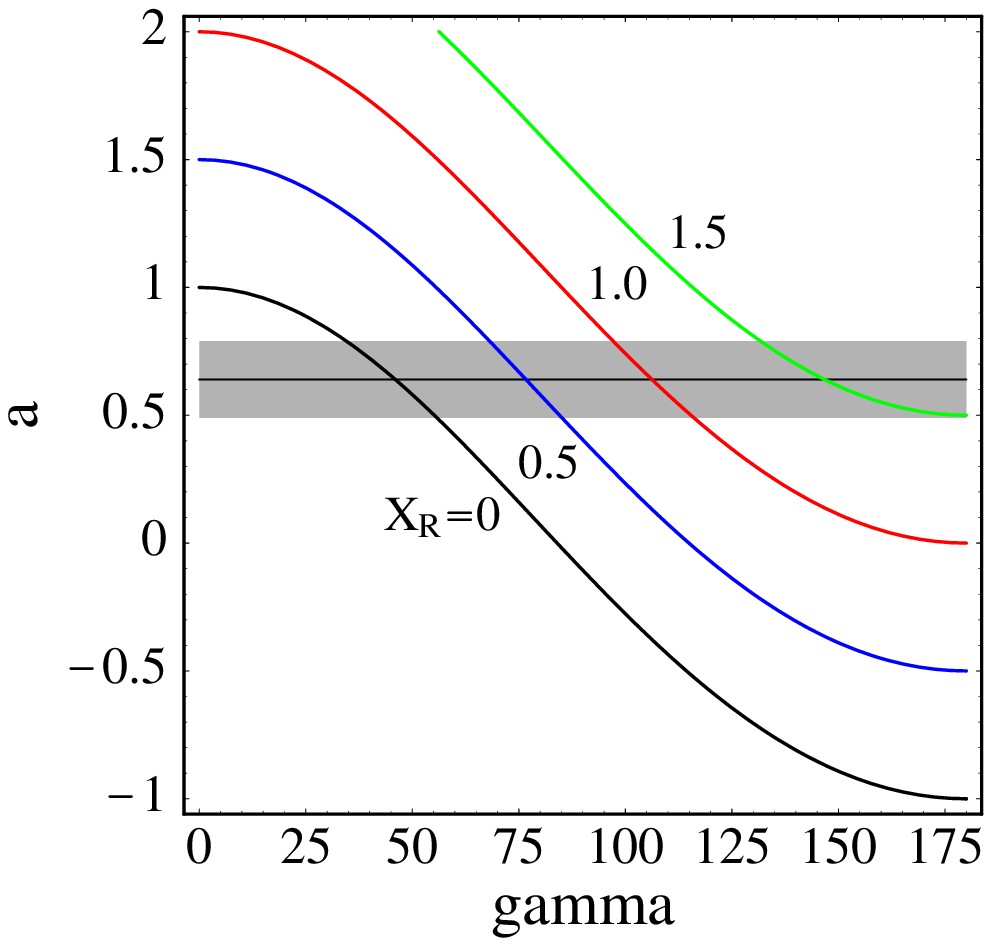,width=7cm} 
\epsfig{file=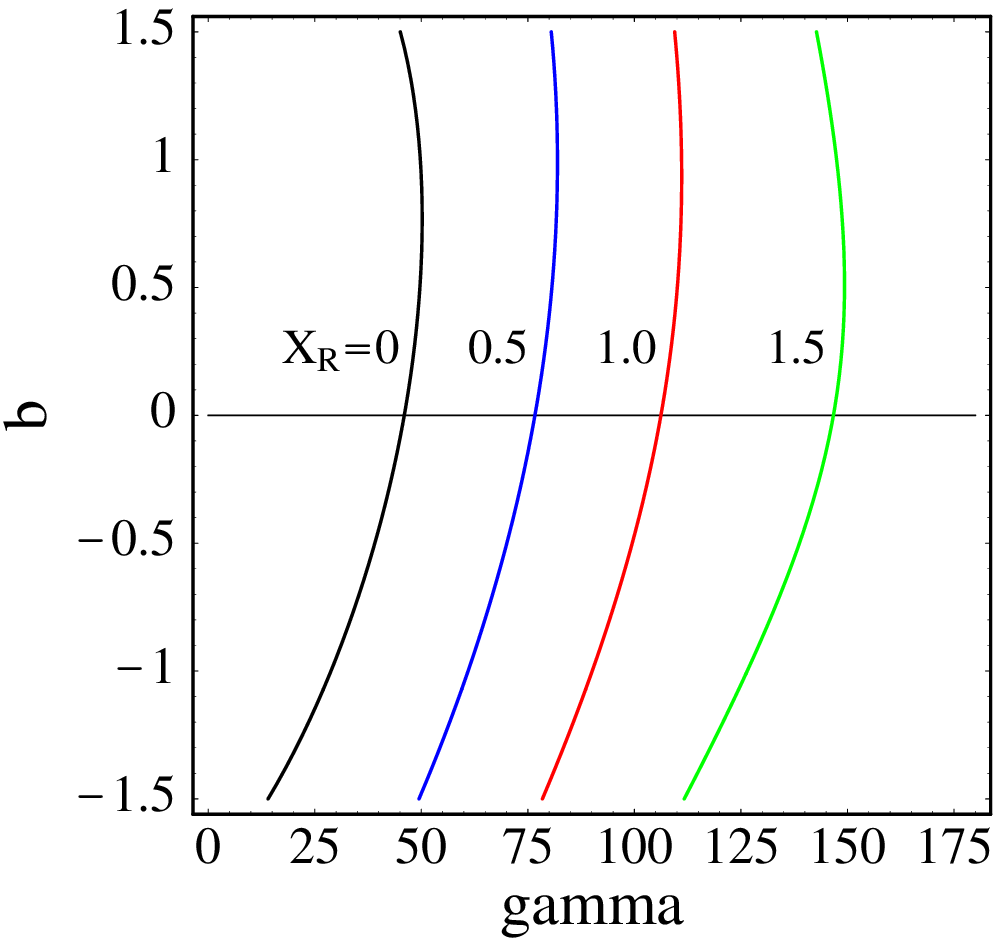,width=7cm} 
\caption{\label{fig:contours}
Contours of constant $X_R$ versus $\gamma$ for different New 
Physics parameters, assuming $\gamma>0$. In each plot one of the 
three parameters $\varphi$, $a$ and $b$ is varied, while the other 
two are kept fixed at their SM values. The horizontal lines 
indicate the SM results.}}

In order to illustrate how big an effect New Physics could have 
on the value of $\gamma$ extracted from $B^\pm\to\pi K$ decays, we 
assume for simplicity that the strong-interaction phase $\phi$ is 
small. We expect that this is indeed a good assumption, since the 
QCD factorization theorem of \cite{fact} predicts that 
$\phi=O(\alpha_s,\Lambda/m_b)$. In this case, a measurement of the 
asymmetry $\widetilde A$ will provide information about $\phi$ 
(and serve as a test of our assumption), whereas $\gamma$ is 
determined by $R_*$ alone. In the context of the SM, the solution
obtained with $\cos\phi\approx 1$ is
\begin{equation}\label{gamextr}
   \cos\gamma_{\pi K} \approx \delta_{\rm EW}
   - \frac{R_*^{-1}-1
           -\bar\varepsilon_{3/2}^2(1-\delta_{\rm EW}^2)}
          {2\bar\varepsilon_{3/2}
           (1+\bar\varepsilon_{3/2}\delta_{\rm EW})} 
   \approx \delta_{\rm EW}
   - \frac{R_*^{-1}-1}{2\bar\varepsilon_{3/2}} \,.
\end{equation}
Given the present uncertainties in $\bar\varepsilon_{3/2}$ and 
$\delta_{\rm EW}$, this result is a good approximation to the
exact solution as long as $|\phi|<25^\circ$. 
Let us now investigate how New Physics may affect the results of
this extraction, neglecting, for the purpose of illustration,
all strong-interaction phases. As in the previous section, we 
focus first on the situation where the New Physics conserves 
isospin symmetry. Inserting for $R_*^{-1}$ in (\ref{gamextr}) the 
expression given in (\ref{Rstphi}), we obtain
\begin{equation}
   \cos\gamma_{\pi K} = \cos(\gamma+\varphi)
   + \delta_{\rm EW}\,(1-\cos\varphi)
   + O(\bar\varepsilon_{3/2}) \,.
\end{equation}
For small values of $\varphi$, the result is simply given by 
$\gamma_{\pi K}\approx\gamma+\varphi$. Therefore, to have a 
significant shift requires having $|\varphi|>25^\circ$, which
corresponds to rather large values $|\rho|>0.5$ and hence an $O(1)$ 
change in the decay amplitudes. The situation is different if the 
New Physics contributions violate isospin symmetry. Consider, for 
simplicity, the case where there are no new CP-violating phases. 
From (\ref{Rsta}) it then follows that
\begin{equation}
   \cos\gamma_{\pi K} = \cos\gamma - a_{\rm NP}
   + O(\bar\varepsilon_{3/2}) \,,
\end{equation}
where
\begin{equation}
   a_{\rm NP}\equiv a - \delta_{\rm EW} \,.
\end{equation}
Now even a moderate New Physics contribution to the electroweak 
penguin coefficients can lead to a large shift in $\gamma$. In 
the most general case, where all New Physics contributions are 
present, we obtain
\begin{equation}
   \cos\gamma_{\pi K} = \cos(\gamma+\varphi) + \delta_{\rm EW} 
   - a\cos\varphi + b\sin\varphi + O(\bar\varepsilon_{3/2}) \,,
\end{equation}
which again allows for large shifts. 

These observations are illustrated in Figure~\ref{fig:contours}, 
where we show contours of constant $X_R$ versus $\gamma$ for 
different values of one of the parameters $\varphi$, $a$ and $b$, 
keeping the other two fixed to their SM values. Without loss of
generality we assume that $\gamma>0$. These plots show that even 
moderate New Physics contributions to the parameter $a$ can induce 
large shifts in $\gamma$. On the other hand, small values of 
$\varphi$ lead to much smaller effects. Likewise, the effects 
induced by the parameter $b$ are much smaller. We stress
that the present central value of $X_R\approx 0.7$ is such that
negative values of $\varphi$, as well as values of $a$ less than 
the SM result $a\approx 0.64$, are disfavored since they would
require values of $\gamma$ exceeding $100^\circ$, in conflict
with the global analysis of the unitarity triangle 
\cite{BaBar,Yuvi}.

\FIGURE{
\epsfig{file=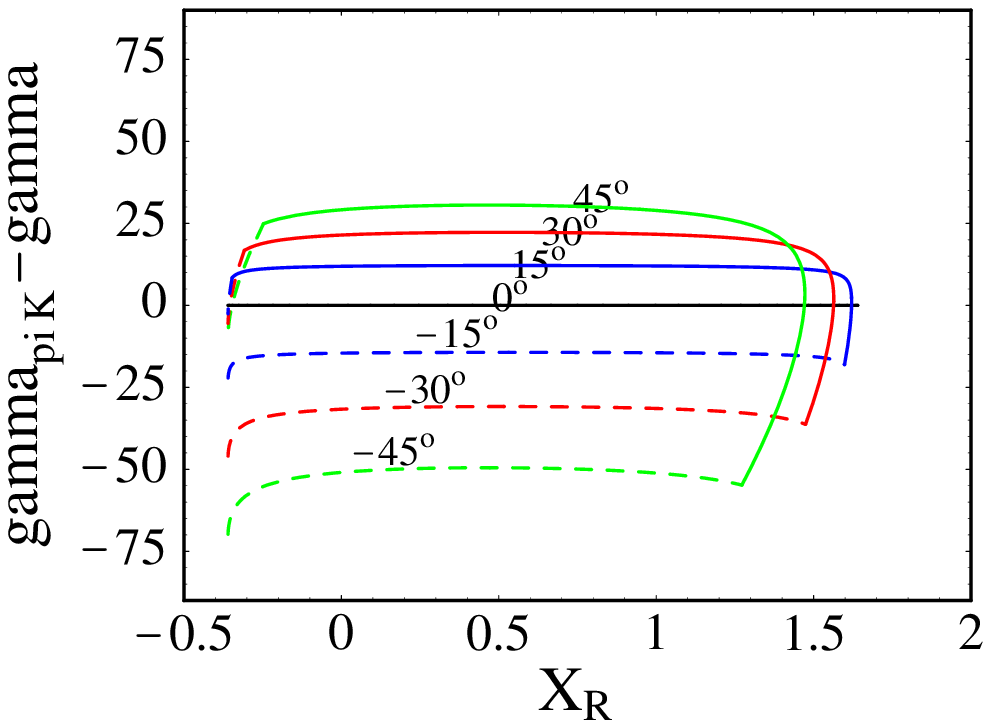,width=7cm}
\epsfig{file=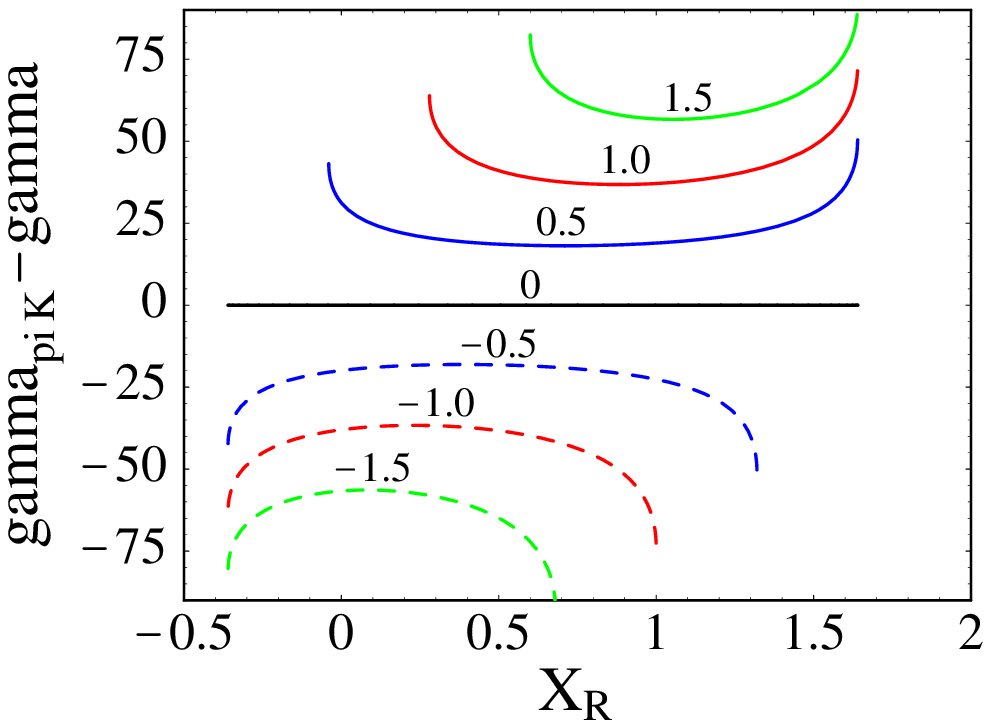,width=7cm} 
\caption{\label{fig:shifts}
Difference $\gamma_{\pi K}-\gamma$ versus $X_R$ for different New 
Physics scenarios, assuming $\gamma>0$. Left: $a=\delta_{\rm EW}$, 
$b=0$ and $\varphi=0$, $\pm 15^\circ$, $\pm 30^\circ$, and 
$\pm 45^\circ$. Right: $\varphi=b=0$ and 
$a_{\rm NP}/\delta_{\rm EW}=0$, $\pm 0.5$, $\pm 1$, and $\pm 1.5$. 
Solid curves refer to positive values of $\varphi$ and 
$a_{\rm NP}$, dashed curves to negative ones.}}

In Figure~\ref{fig:shifts}, we show the difference 
$\gamma_{\pi K}-\gamma$ as a function of $X_R$ for the two cases 
where either $\varphi$ or $a$ are varied with respect to their 
SM values. The dashed curves in the plots refer to negative 
values of $\varphi$ or $a_{\rm NP}$, which are disfavored by
the present value of $X_R$. This implies that isospin-conserving
New Physics can only lead to moderate shifts in $\gamma$, which
reach the $30^\circ$ level for large values $\rho=O(1)$.
Isospin-violating New Physics effects, on the other hand, can
induce very large shifts of $\gamma$ even if they are of 
moderate size. As an example, consider the case 
where a future, precise measurement would yield $X_R=1$. An 
interpretation of this result in the SM would imply a 
relatively large value $\gamma\approx 106^\circ$, which can be 
determined with a theoretical uncertainty of about $10^\circ$ or 
better \cite{NR,me}. Imagine that all other information about 
the unitarity triangle favors a value of $\gamma\approx 75^\circ$, 
again with a small error. To accommodate this difference, one could 
either invoke a large, isospin-conserving but CP-violating New 
Physics contribution such that $\varphi\approx 45^\circ$ 
(corresponding to $\rho\approx 1$), or an isospin-violating 
electroweak penguin contribution such that $a$ is twice as large 
as in the SM. The first solution would imply a New Physics
contribution to the decay amplitudes of order 100\%, whereas the 
second would imply only a small contribution of less than 15\%. 

\subsection{``Wrong kaon'' decays}

So far, when considering $B\to\pi K$ decays we have implicitly 
assumed an underlying quark transition of the form 
$\bar b\to\bar s q\bar q$, in which case the decays with a 
neutral kaon in the final state are $B^+\to\pi^+ K^0$ and 
$B^-\to\pi^-\bar K^0$. Indeed, in the SM this is an excellent 
approximation, because the quark transition 
$\bar b\to\bar d s\bar d$ leading to the ``wrong kaon'' decays 
$B^+\to\pi^+\bar K^0$ and $B^-\to\pi^- K^0$ is highly suppressed. 
However, this may no longer be the case in the presence of New 
Physics.\footnote{For the related decay 
$\bar b\to\bar s d\bar s$, this possibility has been explored in 
\protect\cite{bssd}.}

In practice, only the $B^\pm\to\pi^\pm K_{S,L}$ decay rates can 
be measured. In particular, the CLEO result for $R_*$ quoted in 
(\ref{Rstexp}) really refers to the ratio
\begin{equation}\label{Xwrong}
   R_*^{\rm exp} = 
    \frac{\mbox{B}(B^+\to\pi^+ K_S)+\mbox{B}(B^-\to\pi^- K_S)}
         {\mbox{B}(B^+\to\pi^0 K^+)+\mbox{B}(B^-\to\pi^0 K^-)} 
   \equiv R_*\,(1+|X_{\rm wrong}|^2) \,,
\end{equation}
which differs from $R_*$ by the ``wrong kaon'' contribution
\begin{equation}
   |X_{\rm wrong}|^2 =  
   \frac{\mbox{B}(B^+\to\pi^+\bar K^0)+\mbox{B}(B^-\to\pi^- K^0)}
        {\mbox{B}(B^+\to\pi^+ K^0)+\mbox{B}(B^-\to\pi^-\bar K^0)}
   \,.
\end{equation}
Note that the presence of this contribution could only enhance the 
observed value $R_*^{\rm exp}$ with respect to $R_*$. This 
observation, combined with the fact that $R_*^{\rm exp}$ is not much 
larger than the expected value for $R_*$ in the SM, allows us to put 
bounds on the ``wrong kaon'' contribution in specific New Physics 
models. 

In analogy with (\ref{basis}), there are three operators entering 
the effective Hamiltonian for $\bar b\to\bar d s\bar d$ decays, 
which we define as
\begin{eqnarray}
   O_1^{dd} &=& (\bar b_\alpha d_\alpha)_{V-A}\,
    (\bar s_\beta d_\beta)_{V+A} \,, \nonumber\\
   O_2^{dd} &=& (\bar b_\alpha d_\beta)_{V-A}\,
    (\bar s_\beta d_\alpha)_{V+A} \,, \nonumber\\
   O_3^{dd} &=& (\bar b_\alpha d_\alpha)_{V-A}\,
    (\bar s_\beta d_\beta)_{V-A} \,,
\end{eqnarray}
and similarly there may be operators $\widetilde O_i^{dd}$ of 
opposite chirality. We denote the corresponding Wilson 
coefficients by $c_i^{dd}$ and $\widetilde c_i^{dd}$, respectively. 
The renormalization-group evolution of these coefficients can be 
read off from (\ref{RGE-R}) and (\ref{RGE-L}) by replacing 
$c_1^{\rm EW}\to c_1^{dd}$, $c_2^{\rm EW}\to c_2^{dd}$, and
$c_{3,4}^{\rm EW}\to c_3^{dd}$. Using factorization, and 
normalizing the result to $\bar\varepsilon_{3/2}$, we find
\begin{equation}
   X_{\rm wrong} = \frac{\bar\varepsilon_{3/2}}{|\lambda_u|}
   \left[ \bar c_3^{dd} - \frac23\,\kappa^{3/23}\,\bar c_1^{dd}
   + \frac{3\chi-1}{4}\,\kappa^{30/23} \bigg( \bar c_2^{dd}
   + \frac13\,\bar c_1^{dd} \bigg) \right] \,,
\end{equation}
where $\bar c_i^{dd}\equiv c_i^{dd}(m_W)-\widetilde c_i^{dd}(m_W)$.
Inserting this result into (\ref{Xwrong}), and performing the
evolution to the scale $\mu=m_b$, we obtain
\begin{equation}\label{Xwrongbound}
   |\bar c_3^{dd} - 0.26 \bar c_1^{dd} + 1.40\bar c_2^{dd}|
   \approx \frac{|\lambda_u|}{\bar\varepsilon_{3/2}}
   \sqrt{\frac{R_*^{\rm exp}}{R_*} -1} \,.
\end{equation}
Using the current values of $\lambda_u$, $\bar\varepsilon_{3/2}$ and
$R_*^{\rm exp}$, and taking $R_*>0.45$ corresponding to the smallest
possible value in the SM, we find that the right-hand side is less
than $5.2\times 10^{-3}$ at 90\% confidence level.

\section{Trojan penguins from tree-level processes}
\label{sec:tree}

In this and the following sections, we calculate the matching 
conditions for the penguin coefficients $c_i^q$ and 
$\widetilde c_i^q$ in a variety of extensions of the SM with new 
flavor-violating couplings. We focus first on models where such 
couplings appear at tree level. Loop-mediated New Physics 
contributions will be discussed in Section~\ref{sec:loop}.

\subsection{Flavor-changing $Z$-boson exchange} 

A generic feature of extensions of the SM with extra
nonsequential quarks is the presence of tree-level flavor-changing
couplings of the $Z$ boson, such as a $bsZ$ vertex. A detailed
discussion of such models can be found, e.g., in \cite{bsnn}. 
The relevant terms in the Lagrangian read
\begin{equation}\label{mixcurrent}
   {\cal L}_{\rm FCNC} = -\frac{g}{4\cos\theta_W}\,\sum_{i\neq j} 
   \,\bar d^i\,[\kappa_L^{ij}\gamma^\mu(1-\gamma_5) + 
   \kappa_R^{ij}\gamma^\mu(1+\gamma_5)]\,d^j Z_\mu \,,
\end{equation}
where $i,j$ are generation indices. The quantities $\kappa_L^{bs}$ 
and $\kappa_R^{bs}$ are the two new complex parameters relevant to 
$b\to s$ transitions. Since the flavor-violating interactions are 
small (see below), the flavor-diagonal couplings of the $Z$-boson 
are to leading order the same as in the SM. It follows that at low 
energies the tree-level $Z$ exchange for the decay 
$\bar b\to\bar s q\bar q$ leads to the effective Hamiltonian 
\begin{equation}\label{ZEWP}
   {\cal H}_{\rm eff} = \frac{G_F}{\sqrt2}\,
   [\kappa_L^{bs}(\bar b s)_{V-A} + \kappa_R^{bs}(\bar b s)_{V+A}]
   \,\sum_q\,[ C_L^q(\bar q q)_{V-A} + C_R^q(\bar q q)_{V+A} ] \,,
\end{equation}
where
\begin{eqnarray}
   C_L^u &=& \phantom{-}\frac12 - \frac23 \sin^2\!\theta_W \,,
    \qquad
    C_R^u = -\frac23 \sin^2\!\theta_W \,, \nonumber\\
   C_L^d &=& -\frac12 + \frac13 \sin^2\!\theta_W \,, \qquad
    C_R^d = \phantom{-}\frac13 \sin^2\!\theta_W \,.
\end{eqnarray}
It is straightforward to match this result with the generic form 
of the penguin terms in the effective Hamiltonian, and to deduce
the corresponding values of the Wilson coefficients. The
nonvanishing coefficients at the matching scale are 
$c_1^q=\kappa_L^{bs}\,C_R^q$, $c_3^q=\kappa_L^{bs}\,C_L^q$, 
$\widetilde c_1^q=\kappa_R^{bs}\,C_L^q$, and
$\widetilde c_3^q=\kappa_R^{bs}\,C_R^q$. In the notation of 
(\ref{cdefs}), this implies
\begin{equation}
   c_3^{\rm QCD} = - \frac{\kappa_L^{bs}}{6} \,, \qquad
   \widetilde c_1^{\rm QCD} = - \frac{\kappa_R^{bs}}{6} \,,
   \qquad
   c_1^{\rm QCD} = \widetilde c_3^{\rm QCD} = 0 \,,
\end{equation}
as well as
\begin{eqnarray}
   c_1^{\rm EW} &=& -\kappa_L^{bs}\sin^2\!\theta_W \,, \qquad
    c_3^{\rm EW} = \phantom{-}\kappa_L^{bs}\cos^2\!\theta_W \,,
    \nonumber\\
   \widetilde c_1^{\rm EW} &=& \phantom{-}\kappa_R^{bs}
    \cos^2\!\theta_W \,, \qquad
    \widetilde c_3^{\rm EW} = -\kappa_R^{bs}\sin^2\!\theta_W \,.
\end{eqnarray}
These electroweak penguin coefficients would be of the same order 
as the SM result for $c_3^{\rm EW}$ in (\ref{SMinit}) if 
$|\kappa_{L,R}^{bs}|\approx 5\times 10^{-4}$ which, as we will 
see below, is consistent with experimental bounds. Inserting these 
results into (\ref{abres}) and using $|\lambda_u|\approx 
7.5\times 10^{-4}$ yields
\begin{equation}\label{cute}
   a_{\rm NP}+ib \approx -1.1\times 10^3\,
   (\kappa_L^{bs} + 0.51\kappa_R^{bs}) \,.
\end{equation}
We stress that the simple model considered here is a prototype of 
New Physics models in which the electroweak penguin coefficients 
$c_i^{\rm EW}$ and $\widetilde c_i^{\rm EW}$ are not suppressed 
relative to the QCD penguin coefficients $c_i^{\rm QCD}$ and 
$\widetilde c_i^{\rm QCD}$. This property is in contrast with the 
SM, where electroweak penguins are suppressed by small gauge 
couplings. 

At present, the strongest constraints on $\kappa_L^{bs}$ and 
$\kappa_R^{bs}$ follow from the experimental bound on the 
$B\to X_s\,e^+ e^-$ decay rate. Since this bound lies far above 
the SM prediction for this process, we can neglect the SM 
contribution and write the effective Hamiltonian for this 
process as
\begin{equation}
   {\cal H}_{\rm eff} = \frac{G_F}{\sqrt2}\,
   [\kappa_L^{bs}(\bar bs)_{V-A} + \kappa_R^{bs}(\bar bs)_{V+A}] 
   \left[ C_L^e\,(\bar e e)_{V-A}
   + C_R^e\,(\bar e e)_{V+A} \right] \,,
\end{equation}
where
\begin{equation}
   C_L^e = -\frac12 + \sin^2\!\theta_W \,, \qquad
   C_R^e = \sin^2\!\theta_W \,. 
\end{equation}
It is convenient to normalize the result for the 
$B\to X_s\,e^+ e^-$ decay rate to the semileptonic rate. Then 
many common factors cancel, and we obtain
\begin{equation}
   \frac{\Gamma(B\to X_s\,e^+ e^-)}{\Gamma(B\to X_c\,e^-\bar\nu_e)} 
   = \frac{|\kappa_L^{bs}|^2+|\kappa_R^{bs}|^2}
          {f(m_c/m_b)\,|V_{cb}|^2} 
   \left[ (C_L^e)^2 + (C_R^e)^2 \right] \approx 
   157 \left( |\kappa_L^{bs}|^2 + |\kappa_R^{bs}|^2 \right) \,,
\end{equation}
where we have used $|V_{cb}|\approx 0.04$, and $f(m_c/m_b)\approx 
0.5$ for the phase-space factor in the semileptonic decay. Using 
the upper bound $\mbox{B}(B\to X_s\,e^+ e^-)<5.7\times 10^{-5}$ 
together with $\mbox{B}(B\to X_c\,e^-\bar\nu_e)=0.105$ \cite{PDG} 
yields
\begin{equation}
   \sqrt{|\kappa_L^{bs}|^2 + |\kappa_R^{bs}|^2}
   < 1.9\times 10^{-3} \,.
\end{equation}
Combining this result with (\ref{cute}), we obtain the upper 
bound
\begin{equation}
   \sqrt{a_{\rm NP}^2 + b^2} < 2.0 \,,
\end{equation}
which may be compared with the SM value $a\approx 0.64$. It
follows that tree-level $Z$ exchange with new flavor-violating 
couplings can yield isospin-violating electroweak penguin
effects that are up to a factor 3 larger than in the SM. 

For completeness, we also mention the resulting bound on the New 
Physics parameter $\rho$. Neglecting renormalization-group effects, 
we obtain from (\ref{rhobound}) the estimate
\begin{equation}
   \frac{\rho}{\sqrt{1+\rho^2}} \approx 29\,\Big[
    0.8\,\mbox{Im}\,\kappa_L^{bs} + \mbox{Im}\,\kappa_R^{bs}
    \Big] \quad
   \Rightarrow \quad |\rho| < 0.05 \,, \quad
    |\varphi| < 3^\circ \,,
\end{equation}
indicating that in this model there is no room for large values
of $\rho$.

\subsection{Extended gauge models with a $Z'$ boson}

A new neutral boson $Z'$ with tree-level flavor-changing couplings 
to quarks is a generic property of many models with an extended 
gauge group. The analysis of electroweak penguins in such models 
is very similar to the flavor-changing tree-level $Z$ exchange 
discussed above. For simplicity, we assume no significant mixing 
between the $Z$ and $Z'$ bosons. Then the effective Hamiltonian 
is a simple generalization of (\ref{ZEWP}), i.e.
\begin{eqnarray}\label{Zp-EWP}
   {\cal H}_{\rm eff} &=& \frac{g_{U(1)'}^2}{m_{Z'}^2}
    [ \kappa_L^{\prime bs}(\bar b s)_{V-A}
    + \kappa_R^{\prime bs}(\bar b s)_{V+A} ]\,
   \sum_q\, [ C_L^q(\bar q q)_{V-A} + C_R^q(\bar q q)_{V+A} ] \,,
\end{eqnarray}
where $C_{L,R}^q$ now denote the charges of the quarks under the 
new U$(1)'$ group. Introducing the ratio
\begin{equation}
   \xi = \frac{g_{{\rm U}(1)'}^2}{g^2}\,\frac{m_W^2}{m_{Z'}^2} \,,
\end{equation}
and taking into account that, since we neglect $Z$--$Z'$ mixing, 
the $Z'$ charges are the same for all left-handed fields (in 
particular, $C_L^u=C_L^d$), we find
\begin{eqnarray}
   c_1^{\rm QCD} &=& \xi\kappa_L^{\prime bs}\,
    \frac{C_R^u+2C_R^d}{3} \,, \qquad
    c_3^{\rm QCD} = \xi\kappa_L^{\prime bs} C_L^q \,,
    \nonumber\\
   \widetilde c_1^{\rm QCD} &=& \xi\kappa_R^{\prime bs} C_L^q \,,
    \qquad
   \widetilde c_3^{\rm QCD} = \xi\kappa_R^{\prime bs}\,
    \frac{C_R^u+2C_R^d}{3} \,,
\end{eqnarray}
and
\begin{equation}
   c_1^{\rm EW} = \xi\kappa_L^{\prime bs} (C_R^u-C_R^d) \,, \qquad
   \widetilde c_3^{\rm EW} = \xi\kappa_R^{\prime bs} 
    (C_R^u-C_R^d) \,, \qquad
   c_3^{\rm EW} = \widetilde c_1^{\rm EW} = 0 \,.
\end{equation}
Inserting these results into (\ref{abres}) gives
\begin{equation}
   a_{\rm NP}+ib \approx \frac{\xi(C_R^u-C_R^d)}{|\lambda_u|}\,
   (\kappa_R^{\prime bs}+0.26\kappa_L^{\prime bs}) \,.
\end{equation}

The allowed range for the relevant parameters in $Z'$ extensions 
of the SM is largely model dependent. For example, the bounds 
derived from the upper limit on the $B\to X_s\,l^+ l^-$ branching 
ratio depend on the lepton charges under the new U$(1)'$ gauge 
group. In the so-called ``leptophobic'' $Z'$ models these charges 
are arranged so as to vanish or be very small \cite{leptophobic}. 
Therefore, in general the contributions of the $Z'$ couplings to 
the electroweak penguin operators can be arbitrarily large. In 
fact, the best model-independent bound on these couplings follows 
from the second inequality in (\ref{abbound}), which implies
\begin{equation}
   |\xi (C_R^u-C_R^d)|\,
   |\kappa_R^{\prime bs}+0.26\kappa_L^{\prime bs}|
   < 0.01 \,,
\end{equation}
where we have neglected the small SM contribution. Furthermore,
assuming $C_{L,R}^q=O(1)$ and no cancellations, the bound 
(\ref{rhonumeric}) gives $|\xi\,\mbox{Im}\,
\kappa_{L,R}^{\prime bs}|<O(10^{-2})$. Turning these observations 
around, we conclude that in extended gauge models with 
flavor-changing $Z'$ couplings such that 
$\xi\kappa_{L,R}^{\prime bs}=O(10^{-2})$ there can be very large
New Physics effects in $B\to\pi K$ decays.

\subsection{SUSY models with R-parity violation}

In SUSY models with broken R-parity extra trilinear terms are 
allowed in the superpotential, some of which can give rise to a 
large enhancement of the electroweak penguin coefficients. 
Denoting by $L_L^i$, $Q_L^i$, $u_R^i$ and $d_R^i$ the chiral 
superfields containing, respectively, the left-handed lepton and 
quark doublets, and the right-handed up- and down-type quark 
singlets of the $i$-th generation, these terms read
\begin{equation}\label{Wrpb}
   W = \lambda'_{ijk}\,L_L^i\,Q_L^j\,\bar d_R^k
   + \lambda''_{ijk}\,\bar u_R^i\,\bar d_R^j\,\bar d_R^k \,.
\end{equation}
At low energies, slepton and squark exchange can generate local 
penguin operators. The most general case has been treated 
in \cite{CDK}. For simplicity, we neglect left-right sfermion 
mixing, which is a small effect and does not generate new 
operators. We then find for the coefficients of the various 
penguin operators
\begin{equation}
   \widetilde c_2^u = \sum_{i=1}^3
   \frac{\lambda_{i12}^{\prime *} \lambda'_{i13}}
        {4\sqrt{2} G_F m_{\tilde e_{iL}}^2} \,, \qquad
   - \widetilde c_3^u = \widetilde c_4^u
   = \frac{\lambda_{113}^{\prime\prime *} \lambda''_{112}}
          {2\sqrt{2} G_F m_{\tilde d_{1R}}^2} \,,
\end{equation}
and
\begin{eqnarray}
   c_2^d &=& \sum_{i=1}^3
    \frac{\lambda_{i31}^{\prime *} \lambda'_{i21}}
         {4\sqrt{2} G_F m_{\tilde\nu_i}^2} \,, \qquad
    c_6^d = \sum_{i=1}^3
    \frac{\lambda_{i32}^{\prime *} \lambda'_{i11}}
         {4\sqrt{2} G_F m_{\tilde\nu_i}^2} \,, \nonumber \\
   \widetilde c_2^d &=& \sum_{i=1}^3
    \frac{\lambda_{i12}^{\prime *} \lambda'_{i13}}
         {4\sqrt{2} G_F m_{\tilde\nu_i}^2} \,, \qquad
    \widetilde c_6^d = \sum_{i=1}^3
    \frac{\lambda_{i11}^{\prime *} \lambda'_{i23}}
         {4\sqrt{2} G_F m_{\tilde\nu_i}^2} \,, \nonumber\\
   - \widetilde c_3^d &=& \widetilde c_4^d = \sum_{i=1}^3
    \frac{\lambda_{i13}^{\prime\prime *} \lambda''_{i12}}
         {2\sqrt{2} G_F m_{\tilde u_{iR}}^2} \,.
\end{eqnarray}
Previous authors have investigated bounds on some of these R-parity 
violating couplings in the context of nonleptonic $B$ decays 
\cite{CRS,BD}. However, in these studies model-dependent 
predictions for the overall penguin amplitude $P$ in (\ref{ampls}) 
are employed. The only significant bound which has an impact on our
analysis comes from a combination of constraints derived from limits 
on double nucleon decay into two kaons and neutron--antineutron 
oscillations, yielding $|\lambda_{113}^{\prime\prime *} 
\lambda''_{112}|<10^{-9}$ \cite{CRS}. Therefore, is it safe to 
neglect $\widetilde c_3^u$ and $\widetilde c_4^u$.

Not all of the above coefficients contribute to the 
isospin-violating terms parame\-trized by $a$ and $b$. Up to a 
small SU(2)$_L$ breaking in the slepton and sneutrino masses we 
find $\widetilde c_2^u=\widetilde c_2^d$, and thus $\widetilde 
c_2^{\rm EW}\approx 0$. Moreover, in (\ref{abres}) only the sum 
$\bar c_3^{\rm EW}+\bar c_4^{\rm EW}$ contributes, which vanishes 
since $\widetilde c_3^q=-\widetilde c_4^q$. It follows that
\begin{equation}
   a_{\rm NP}+ib \approx 2.83\times 10^3\,
   \sum_{i=1}^3\,\frac{(100\,\mbox{GeV})^2}{m_{\tilde\nu_i}^2}\, 
   \Big( {\lambda_{i31}^{\prime *} \lambda'_{i21}}
    + \lambda_{i32}^{\prime *} \lambda'_{i11} 
    - \lambda_{i11}^{\prime *} \lambda'_{i23} \Big) \,.
\end{equation}
The result for the parameter $\rho$ is more complicated. Setting
for simplicity all sfermion masses equal, we find from 
(\ref{rhobound})
\begin{eqnarray}
   \frac{\rho}{\sqrt{1+\rho^2}} &\approx& 106\,
    \frac{(100\,\mbox{GeV})^2}{m_{\tilde f}^2} \nonumber\\
   &\times& \sum_{i=1}^3\,\mbox{Im} \Big[
    4 \lambda_{i13}^{\prime\prime *} \lambda''_{i12}
    + 3\chi(\lambda_{i12}^{\prime *} \lambda'_{i13}
            - \lambda_{i31}^{\prime *} \lambda'_{i21})
    + \lambda_{i32}^{\prime *} \lambda'_{i11}
    - \lambda_{i11}^{\prime *} \lambda'_{i23}
    \Big] \,. \qquad
\end{eqnarray}
Using the results derived in Section~\ref{sec:BpiK}, we can obtain
bounds on several of the R-parity violating couplings. Assuming 
that only one combination of couplings is dominant, neglecting the 
SM contribution, and using a common sfermion reference mass of 
$100\,$GeV, we find from (\ref{abbound}) that at 90\% confidence 
level
\begin{eqnarray}\label{bounds1}
   \Big| \sum_{i=1}^3\,\lambda_{i31}^{\prime *} \lambda'_{i21}
   \Big| &<& 4.9\times 10^{-3} \,, \nonumber\\ 
   \Big| \sum_{i=1}^3\,\lambda_{i32}^{\prime *} \lambda'_{i11}|
   \Big| &<& 4.9\times 10^{-3} \,, \nonumber\\ 
   \Big| \sum_{i=1}^3\,\lambda_{i11}^{\prime *} \lambda'_{i23}|
   \Big| &<& 4.9\times 10^{-3} \,.
\end{eqnarray}
In addition, from (\ref{rhonumeric}) we obtain
\begin{eqnarray}\label{bounds2}
   \Big| \sum_{i=1}^3\,\mbox{Im}\,(\lambda_{i13}^{\prime\prime *}
   \lambda''_{i12}) \Big| &<& 3.7\times 10^{-3} \,, \nonumber \\
   \Big| \sum_{i=1}^3\,\mbox{Im}\,(\lambda_{i31}^{\prime *}
   \lambda'_{i21}) \Big| &<& 4.1\times 10^{-3} \,, \nonumber \\
   \Big| \sum_{i=1}^3\,\mbox{Im}\,(\lambda_{i12}^{\prime *}
   \lambda'_{i13}) \Big| &<& 4.1\times 10^{-3} \,, 
\end{eqnarray}
where we do not present bounds on the imaginary parts of couplings 
which are weaker than the corresponding bounds on the absolute 
values in (\ref{bounds1}). 

It is interesting that SUSY models with R-parity violation provide
an example of scenarios in which $\bar b\to\bar d s\bar d$ 
transitions may not be suppressed relative to 
$\bar b\to\bar s d\bar d$ transitions. As pointed out in 
Section~\ref{sec:BpiK}, this can lead to potentially large ``wrong 
kaon'' decays of the type $B^+\to\pi^+\bar K^0$ and 
$B^-\to\pi^- K^0$. In the model considered here, the only 
nonvanishing coefficients are
\begin{equation}
   c_2^{dd} = \sum_{i=1}^3 
   \frac{\lambda_{i31}^{\prime *} \lambda'_{i12}}
        {4\sqrt{2} G_F m_{\tilde\nu_i}^2} \,, \qquad
  \widetilde c_2^{dd} = \sum_{i=1}^3 
   \frac{\lambda_{i21}^{\prime *} \lambda'_{i13}}
        {4\sqrt{2} G_F m_{\tilde\nu_i}^2} \,.
\end{equation}
The result (\ref{Xwrongbound}) can be used to obtain the bounds
\begin{equation}\label{bounds3}
   |\lambda_{i31}^{\prime *} \lambda'_{i12}|
   < 3.4\times 10^{-3} \,, \qquad 
   |\lambda_{i21}^{\prime *} \lambda'_{i13}|
   < 3.4\times 10^{-3} \,, 
\end{equation}
again at 90\% confidence level.

Our bounds in (\ref{bounds1}) and (\ref{bounds2}) are stronger 
than the ones discussed in the literature \cite{CRS,BD} and refer 
to a larger number of $R$-parity violating couplings. Most 
importantly, however, they are affected by much smaller hadronic 
uncertainties. The bounds in (\ref{bounds3}), on the other hand, 
are weaker than constraints derived from $B$--$\bar B$ and 
$K$--$\bar K$ mixing \cite{Bhat}.

\section{Trojan penguins from loop processes}
\label{sec:loop}

Having considered in the previous section some specific models
with tree-level FCNC couplings, we now explore extensions of the
SM in which new contributions to the electroweak and QCD penguin 
operators arise at one-loop order. In particular, we study in
detail the structure of electroweak penguins in SUSY models,
where isospin-violating $\bar b\to\bar s q\bar q$ transitions
can arise due to strong-interaction gluino box diagrams. This
provides another realization of models in which electroweak
penguins are not suppressed relative to QCD penguins. For 
completeness, we also discuss two-Higgs--doublet models 
and models with anomalous gauge-boson couplings. They are simple 
since there are no new CP-violating phases, so only the 
parameter $a$ can receive a New Physics contribution. However, 
there is no parametrical enhancement of the electroweak penguins 
relative to the SM, and thus the New Physics contributions tend 
to be small.

\subsection{SUSY models}

In SUSY extensions of the SM with conserved R-parity, the 
potentially most important contributions to the Wilson 
coefficients of the penguin operators in the effective 
Hamiltonian (\ref{Hpeng}) arise from strong-interaction penguin 
and box diagrams with gluino--squark loops. They can contribute 
to FCNC processes because the gluinos have flavor-changing 
couplings to the quark and squark mass eigenstates. Provided 
there is a significant mass splitting between 
the right-handed up and down squarks, gluino box diagrams are also 
the most important source of isospin violation in SUSY models 
\cite{epspr}. The corresponding contributions to the electroweak 
penguin coefficients are then much more important than other SUSY 
contributions from photon or $Z$ penguins usually discussed in the 
literature. In fact, in such a scenario SUSY contributions to the 
coefficients of the electroweak and QCD penguin operators are of 
the same order and scale like $\alpha_s^2/m_{\rm SUSY}^2$, where 
$m_{\rm SUSY}$ is a generic mass of the superparticles. Whereas the 
QCD penguin contributions are typically smaller than in the SM, the 
electroweak penguin contributions can be important, since their 
scaling relative to the SM coefficients is controlled by the ratio 
$(\alpha_s/\alpha)(m_W^2/m_{\rm SUSY}^2)\sim 1$. In our analysis we 
will consider only these potentially large gluino box and penguin 
contributions and neglect a multitude of other SUSY diagrams, which 
are parametrically suppressed by small electroweak gauge couplings. 
The latter include photon or $Z$ penguins with gluino--squark, 
chargino--squark, neutralino--squark, or charged-Higgs--quark loops, 
and various box diagrams containing at least one chargino or 
neutralino. We have calculated all of these diagrams and, for 
generic regions in SUSY parameter space, have found their 
contributions to be largely suppressed relative to the pure gluino 
diagrams.

Large SUSY contributions to the penguin operators via gluino 
loops require near maximal mixing between the strange and 
bottom squarks, so that the squark mass-insertion approximation 
is not valid. We therefore present our results using the general 
vertex-mixing method, summing over diagrams with different squark 
mass-eigenstates in the loops \cite{berto91,cho}. We denote by 
$\Gamma^{D_L}$ the rotation matrices relating the left-handed 
down-squark interaction states in the quark mass-eigenbasis, 
$\tilde q_L^I$ ($q=d,s,b$), to the squark mass eigenstates, 
$\tilde d_i$ ($i=1,\dots,6$), such that 
$\tilde q_L^I=(\Gamma^{D_L}_{iq})^*\,\tilde{d}_i$, with obvious 
generalizations for the up- and right-handed squarks. In addition, 
$x_{\tilde q_i\tilde g}\equiv m_{\tilde q_i}^2/m_{\tilde g}^2$, 
where $m_{\tilde q_i}$ is the mass of the $i$-th down ($q=d$) or 
up ($q=u$) squark mass eigenstate. In the operator basis of 
(\ref{basis}), we obtain for the gluino box contributions to the 
Wilson coefficients at the SUSY matching scale
\begin{eqnarray}\label{gluinobox}
   c_{1,{\rm box}}^u
   &=& \frac{\alpha_s^2}{2\sqrt2 G_F m_{\tilde g}^2}
    \left(\Gamma^{D_L}_{ib}\right)^* \Gamma^{D_L}_{is}
    \left(\Gamma^{U_R}_{ju}\right)^* \Gamma^{U_R}_{ju} 
    \left[ \frac{1}{18}\,
     F(x_{\tilde d_i\tilde g},x_{\tilde u_j\tilde g})
     - \frac{5}{18}\,
     G(x_{\tilde d_i\tilde g},x_{\tilde u_j\tilde g}) \right] ,
    \nonumber\\
   c_{1,{\rm box}}^d
   &=& \frac{\alpha_s^2}{2\sqrt2 G_F m_{\tilde g}^2}
    \left(\Gamma^{D_L}_{ib}\right)^* \Gamma^{D_L}_{is}
    \left(\Gamma^{D_R}_{jd}\right)^* \Gamma^{D_R}_{jd} 
    \left[ \frac{1}{18}\,
     F(x_{\tilde d_i\tilde g},x_{\tilde d_j\tilde g})
     - \frac{5}{18}\,
     G(x_{\tilde d_i\tilde g},x_{\tilde d_j\tilde g}) \right] ,
    \nonumber\\
   c_{2,{\rm box}}^u
   &=& \frac{\alpha_s^2}{2\sqrt2 G_F m_{\tilde g}^2}
    \left(\Gamma^{D_L}_{ib}\right)^* \Gamma^{D_L}_{is}
    \left(\Gamma^{U_R}_{ju}\right)^* \Gamma^{U_R}_{ju} 
    \left[ \frac76\,
     F(x_{\tilde d_i\tilde g},x_{\tilde u_j\tilde g})
     + \frac16\,
     G(x_{\tilde d_i\tilde g},x_{\tilde u_j\tilde g}) \right] ,
    \nonumber\\
   c_{2,{\rm box}}^d
   &=& \frac{\alpha_s^2}{2\sqrt2 G_F m_{\tilde g}^2}
    \left(\Gamma^{D_L}_{ib}\right)^* \Gamma^{D_L}_{is}
    \left(\Gamma^{D_R}_{jd}\right)^* \Gamma^{D_R}_{jd} 
    \left[ \frac76\,
     F(x_{\tilde d_i\tilde g},x_{\tilde d_j\tilde g})
     + \frac16\,
     G(x_{\tilde d_i\tilde g},x_{\tilde d_j\tilde g}) \right] ,
    \nonumber\\
   c_{3,{\rm box}}^u
   &=& \frac{\alpha_s^2}{2\sqrt2 G_F m_{\tilde g}^2}
    \left(\Gamma^{D_L}_{ib}\right)^* \Gamma^{D_L}_{is}
    \left(\Gamma^{U_L}_{ju}\right)^* \Gamma^{U_L}_{ju} 
    \left[ -\frac59\,
     F(x_{\tilde d_i\tilde g},x_{\tilde u_j\tilde g})
     + \frac{1}{36}\,
     G(x_{\tilde d_i\tilde g},x_{\tilde u_j\tilde g}) \right] ,
    \nonumber\\
   c_{3,{\rm box}}^d
   &=& \frac{\alpha_s^2}{2\sqrt2 G_F m_{\tilde g}^2}
    \Bigg\{ \left(\Gamma^{D_L}_{ib}\right)^* \Gamma^{D_L}_{is}
    \left(\Gamma^{D_L}_{jd}\right)^* \Gamma^{D_L}_{jd} 
    \left[ -\frac59\,
     F(x_{\tilde d_i\tilde g},x_{\tilde d_j\tilde g})
     + \frac{1}{36}\,
     G(x_{\tilde d_i\tilde g},x_{\tilde d_j\tilde g}) \right]
    \nonumber\\
   &&\hspace{1.8cm}\mbox{}+ 
    \left(\Gamma^{D_L}_{ib}\right)^* \Gamma^{D_L}_{js}
    \left(\Gamma^{D_L}_{jd}\right)^* \Gamma^{D_L}_{id} 
    \left[ \frac13\,
     F(x_{\tilde d_i\tilde g},x_{\tilde d_j\tilde g})
     - \frac{7}{12}\,
     G(x_{\tilde d_i\tilde g},x_{\tilde d_j\tilde g}) \right]
    \Bigg\} , \nonumber\\
   c_{4,{\rm box}}^u
   &=& \frac{\alpha_s^2}{2\sqrt2 G_F m_{\tilde g}^2}
    \left(\Gamma^{D_L}_{ib}\right)^* \Gamma^{D_L}_{is}
    \left(\Gamma^{U_L}_{ju}\right)^* \Gamma^{U_L}_{ju} 
    \left[ \frac13\,
     F(x_{\tilde d_i\tilde g},x_{\tilde u_j\tilde g})
     - \frac{7}{12}\,
     G(x_{\tilde d_i\tilde g},x_{\tilde u_j\tilde g}) \right] ,
    \nonumber\\
   c_{4,{\rm box}}^d
   &=& \frac{\alpha_s^2}{2\sqrt2 G_F m_{\tilde g}^2}
    \Bigg\{ \left(\Gamma^{D_L}_{ib}\right)^* \Gamma^{D_L}_{is}
    \left(\Gamma^{D_L}_{jd}\right)^* \Gamma^{D_L}_{jd} 
    \left[ \frac13\,
     F(x_{\tilde d_i\tilde g},x_{\tilde d_j\tilde g})
     - \frac{7}{12}\,
     G(x_{\tilde d_i\tilde g},x_{\tilde d_j\tilde g}) \right]
    \nonumber\\
   &&\hspace{1.8cm}\mbox{}+ 
    \left(\Gamma^{D_L}_{ib}\right)^* \Gamma^{D_L}_{js}
    \left(\Gamma^{D_L}_{jd}\right)^* \Gamma^{D_L}_{id} 
    \left[ -\frac59\,
     F(x_{\tilde d_i\tilde g},x_{\tilde d_j\tilde g})
     + \frac{1}{36}\,
     G(x_{\tilde d_i\tilde g},x_{\tilde d_j\tilde g}) \right]
    \Bigg\} , \nonumber\\
   c_{5,{\rm box}}^d
   &=& \frac{\alpha_s^2}{2\sqrt2 G_F m_{\tilde g}^2}
    \left(\Gamma^{D_L}_{ib}\right)^* \Gamma^{D_R}_{js}
    \left(\Gamma^{D_R}_{jd}\right)^* \Gamma^{D_L}_{id} 
    \left[ \frac{1}{18}\,
     F(x_{\tilde d_i\tilde g},x_{\tilde d_j\tilde g})  
     - \frac{5}{18}\,
     G(x_{\tilde d_i\tilde g},x_{\tilde d_j\tilde g}) \right] , 
    \nonumber\\
   c_{6,{\rm box}}^d
   &=& \frac{\alpha_s^2}{2\sqrt2 G_F m_{\tilde g}^2}
    \left(\Gamma^{D_L}_{ib}\right)^* \Gamma^{D_R}_{js}
    \left(\Gamma^{D_R}_{jd}\right)^* \Gamma^{D_L}_{id} 
    \left[ \frac76\,
     F(x_{\tilde d_i\tilde g},x_{\tilde d_j\tilde g})  
     + \frac16\,
     G(x_{\tilde d_i\tilde g},x_{\tilde d_j\tilde g}) \right] , 
\end{eqnarray}
where repeated indices are summed over, and $c_{5,{\rm box}}^u
=c_{6,{\rm box}}^u=0$. The functions $F(x,y)$ and $G(x,y)$ are 
given by \cite{berto91,cho} 
\begin{eqnarray}
   F(x,y) &=& -\frac{x\ln x}{(x-y)(x-1)^2}
    - \frac{y\ln y}{(y-x)(y-1)^2} - \frac{1}{(x-1)(y-1)} \,,
    \nonumber\\
   G(x,y) &=& \frac{x^2\ln x}{(x-y)(x-1)^2}
    + \frac{y^2\ln y}{(y-x)(y-1)^2} + \frac{1}{(x-1)(y-1)} \,. 
\end{eqnarray}
The corresponding expressions for the coefficients 
$\widetilde c_i^q$ of the opposite-chirality operators are 
obtained via the exchange $L\leftrightarrow R$ in the expressions 
for $c_i^q$. In practice, $c_{5,{\rm box}}^d$, $c_{6,{\rm box}}^d$ 
and the second terms in $c_{3,{\rm box}}^d$, $c_{4,{\rm box}}^{d}$
(as well as the corresponding terms in the coefficients of the 
opposite-chirality operators) can be neglected due to 
$B_d$--$\bar B_d$ and $K$--$\bar K$ mixing constraints on the 
off-diagonal (1-3) and (1-2) entries of the $\Gamma^{D_L}$ and 
$\Gamma^{D_R}$ matrices. Gluino box diagrams which in the 
mass-insertion approximation would contain left--right squark-mass 
insertions can also be neglected and have not been included in 
(\ref{gluinobox}). Specifically, these graphs would contain the 
mass insertions $\delta m^2_{\tilde s_L\tilde b_R}$, 
$\delta m^2_{\tilde s_R\tilde b_L}$, 
$\delta m^2_{\tilde d_L\tilde b_R}$  
or $\delta m^2_{\tilde d_R\tilde b_L}$, whose magnitudes are 
tightly constrained by the experimental value of the 
$B\to X_{s,d}\,\gamma$ branching ratio \cite{GGMS96}. Further 
suppression of such graphs can be expected on theoretical grounds 
since the remaining left--right squark-mass insertions they would 
contain are suppressed by light quark masses in general 
supergravity theories \cite{Louis,lostAlex} and in SUSY theories 
of flavor \cite{Dine,nirseiberg,susyflavor}. 

In addition to the box contributions, the QCD penguin coefficients 
also receive contributions from gluon penguin diagrams containing 
gluino--squark loops. These are given at the SUSY matching scale by 
\cite{barbi,abel}
\begin{eqnarray}\label{gluinoQCD}
   c_{1,{\rm peng}}^q &=& c_{3,{\rm peng}}^q
    = -\frac{c_{2,{\rm peng}}^q}{3}
    = -\frac{c_{4,{\rm peng}}^q}{3} \nonumber\\
   &=& \frac{\alpha_s^2}{2\sqrt2 G_F m_{\tilde g}^2}
    \left( \Gamma^{D_L}_{ib} \right)^* \Gamma^{D_L}_{is} 
    \left[ \frac12\, A(x_{\tilde d_i\tilde g})
    + \frac29\,B(x_{\tilde d_i\tilde g}) \right] ,
\end{eqnarray}
where
\begin{eqnarray}
   A(x) &=& \frac{1}{2(1-x)} + \frac{(1+2x)\ln x}{6(1-x)^2} \,,
    \nonumber \\
   B(x) &=& - \frac{11-7x+2x^2}{18(1-x)^3}
    - \frac{\ln x}{3(1-x)^4} \,.
\end{eqnarray}
The opposite-chirality contributions are again obtained via the 
substitution $L\rightarrow R$. 
  
FCNC constraints on the off-diagonal entries of the squark mass 
matrix allow for a simple parametrization of the gluino box and 
penguin contributions to $\bar b\to\bar s u\bar u$ and 
$\bar b\to\bar s d\bar d$ transitions, up to small corrections 
which have a negligible impact on the resulting Wilson 
coefficients. Let us first consider the down-squark sector. 
Constraints from $B_d$--$\bar B_d$ and $K$--$\bar K$ mixing imply 
that, to good approximation, the down squark is decoupled from 
the strange and bottom squarks. We also neglect the left--right 
down-squark submatrix, since even in the most general case of 
supergravity theories with arbitrary K\"ahler potential its 
entries are much smaller than the typical squark-mass squared 
\cite{Louis,lostAlex},\footnote{The largest entries in the absence 
of flavor symmetries are suppressed by a factor 
$m_b/m_{\rm SUSY}$.}
and in SUSY theories of flavor its entries are even more 
suppressed \cite{susyflavor}. The above simplifications essentially 
give three ``left-handed'' and three ``right-handed'' down-squark 
mass eigenstates, obtained by diagonalizing the left--left and 
right--right squark submatrices. Specifically, in the physical 
down-quark basis $(d_L,s_L,b_L)$ the left--left submatrix takes 
the form
\begin{equation}
   M_{d,LL}^2 \simeq \left( \begin{array}{ccc}
   m_{11}^2 & 0 & 0 \\
   0 & m_{22}^2 & m^2_{23} \\
   0 & (m_{23}^2)^* & m_{33}^2
   \end{array} \right) \,. 
\end{equation}
Let us denote the left-handed mass eigenstates by $\tilde q_L$ 
($q=d,s,b$) and their masses by $m_{\tilde q_L}^2$, making the 
identification $m_{\tilde d_L}^2=m_{11}^2$. Then the left-handed 
squark mass-eigenstates take the form
\begin{equation}
   \tilde d_L\equiv \tilde d_1 = \left( \begin{array}{c}
    1 \\ 0 \\ 0 \end{array} \right) \,, \quad
   \tilde s_L\equiv \tilde d_2 = \left( \begin{array}{c}
    0 \\ \cos\theta_L \\ -\sin\theta_L\,e^{-i\delta_L}
    \end{array} \right) \,, \quad
   \tilde b_L\equiv \tilde d_3 = \left( \begin{array}{c}
    0 \\ \sin\theta_L\,e^{i\delta_L} \\ \cos\theta_L
    \end{array} \right) \,,
\end{equation} 
where $\delta_L$ is a new CP-violating phase. We take $|\theta_L|
\le 45^\circ$, so that the squark mass-eigenstate $\tilde s_L$ is 
more closely aligned with the $s$ quark, and $\tilde b_L$ with 
the $b$ quark. In the case of the box graphs we also need to 
consider the up-squark sector. $D$--$\bar D$ mixing bounds 
\cite{nirseiberg} imply that, to good approximation, the up 
squark is decoupled from the charm squark in the sense that 
including the phenomenologically allowed mixing between the two 
will lead to negligible modifications of the Wilson coefficients. 
Without loss of generality, we can also ignore mixing between the 
up and top squarks, which is a good approximation in SUSY theories 
of flavor.\footnote{Although up--top squark mixing can be large in 
supergravity theories with arbitrary K\"ahler potential, it would 
not modify our conclusions qualitatively. Furthermore, a model 
which would admit large up--top squark mixing but satisfy the 
$D$--$\bar D$ constraint on up--charm squark mixing would have to 
be very contrived.} 
Finally, as before we can neglect the mixing between the left- and 
right-handed up squarks. 

Taken together, the above approximations imply
\begin{eqnarray}\label{simpleap}
   \left(\Gamma^{D_L}_{ib}\right)^* \Gamma^{D_L}_{is}
   &\simeq& \frac12\sin2\theta_L\,e^{i\delta_L}\,
    (\delta_{\tilde d_i\tilde b_L}
    - \delta_{\tilde d_i\tilde s_L}) \,, \nonumber\\
   \left(\Gamma^{D_L}_{ib}\right)^* \Gamma^{D_L}_{id}
   &\simeq& 0 \,, \nonumber\\
   \left(\Gamma^{D_L}_{jd}\right)^* \Gamma^{D_L}_{jd}
   &\simeq& \delta_{\tilde d_j\tilde d_L} \,, \qquad
    \left(\Gamma^{U_L}_{ju}\right)^* \Gamma^{U_L}_{ju}
    \simeq \delta_{\tilde u_j\tilde u_L} \,.
\end{eqnarray}
The diagonalization in the right-handed sector proceeds in a 
similar way, leading to mass eigenstates parameterized by a mixing 
angle $\theta_R$ and a weak phase $\delta_R$. 
  
It is now straightforward to reexpress the gluino box and penguin  
contributions in (\ref{gluinobox}) and (\ref{gluinoQCD}) in terms 
of our parametrization. The combined results for the coefficients 
$c_i^q$ read ($q=u,d$) 
\begin{eqnarray}\label{gluinoboxparam}
   c_1^q &=& \frac{\alpha_s^2\sin2\theta_L\,e^{i\delta_L}}
          {4\sqrt2 G_F m_{\tilde g}^2} \left[
    \frac{1}{18}\,F(x_{\tilde b_L\tilde g},x_{\tilde q_R\tilde g}) 
    - \frac{5}{18}\,G(x_{\tilde b_L\tilde g},x_{\tilde q_R\tilde g}) 
    + \frac12\,A(x_{\tilde b_L\tilde g}) 
    + \frac{2}{9}\,B(x_{\tilde b_L\tilde g}) \right] \nonumber\\
   &&\mbox{}- (x_{\tilde b_L\tilde g}\to x_{\tilde s_L\tilde g})
    \,, \nonumber\\
   c_2^q &=& \frac{\alpha_s^2\sin2\theta_L\,e^{i\delta_L}}
          {4\sqrt2 G_F m_{\tilde g}^2} \left[
    \frac76\,F(x_{\tilde b_L\tilde g},x_{\tilde q_R\tilde g}) 
    + \frac16\,G(x_{\tilde b_L\tilde g},x_{\tilde q_R\tilde g}) 
    - \frac32\,A(x_{\tilde b_L\tilde g}) 
    - \frac{2}{3}\,B(x_{\tilde b_L\tilde g}) \right] \nonumber\\
   &&\mbox{}- (x_{\tilde b_L\tilde g}\to x_{\tilde s_L\tilde g})
    \,, \nonumber\\
   c_3^q &=& \frac{\alpha_s^2\sin2\theta_L\,e^{i\delta_L}}
          {4\sqrt2 G_F m_{\tilde g}^2} \left[
    - \frac59\,F(x_{\tilde b_L\tilde g},x_{\tilde q_L\tilde g}) 
    + \frac{1}{36}\,G(x_{\tilde b_L\tilde g},x_{\tilde q_L\tilde g}) 
    + \frac12\,A(x_{\tilde b_L\tilde g}) 
    + \frac{2}{9}\,B(x_{\tilde b_L\tilde g}) \right] \nonumber\\
   &&\mbox{}- (x_{\tilde b_L\tilde g}\to x_{\tilde s_L\tilde g})
    \,, \nonumber\\
   c_4^q &=& \frac{\alpha_s^2\sin2\theta_L\,e^{i\delta_L}}
          {4\sqrt2 G_F m_{\tilde g}^2} \left[
    \frac13\,F(x_{\tilde b_L\tilde g},x_{\tilde q_L\tilde g}) 
    - \frac{7}{12}\,G(x_{\tilde b_L\tilde g},x_{\tilde q_L\tilde g}) 
    - \frac32\,A(x_{\tilde b_L\tilde g}) 
    - \frac{2}{3}\,B(x_{\tilde b_L\tilde g}) \right] \nonumber\\
   &&\mbox{}- (x_{\tilde b_L\tilde g}\to x_{\tilde s_L\tilde g})
    \,.
\end{eqnarray}
The coefficients $\widetilde c_i^q$ are obtained by substituting 
$L\leftrightarrow R$ above. The coefficients $c_{5,6}^q$ and 
$\widetilde c_{5,6}^q$ vanish in the approximation 
(\ref{simpleap}).

Let us identify those regions of SUSY parameter space which can 
give large contributions to the Wilson coefficients $c_i^q$ and
$\widetilde c_i^q$. It turns out that a small gluino mass is  
favored for all of the coefficients. Large contributions also 
require $m_{\tilde s_{L,R}}^2\gg m_{\tilde b_{L,R}}^2$ and small 
$m_{\tilde b_{L,R}}^2$, or vice versa. Both options, 
$m_{\tilde s_{L,R}}^2\gg m_{\tilde b_{L,R}}^2$ or
$m_{\tilde b_{L,R}}^2\gg m_{\tilde s_{L,R}}^2$, are equivalent 
as far as the magnitudes of the new contributions to the Wilson 
coefficients are concerned. Perhaps the first option is 
more attractive given that constraints from $K$--$\bar K$ mixing 
are more stringent than those from $B$--$\bar B$ mixing. In 
addition, in models where SUSY is broken at high energies, e.g., 
at the grand-unified theory (GUT) or Planck scales, 
renormalization tends to make the third-generation squarks the 
lightest at the weak scale. Large contributions also require near 
maximal mixing between the strange and bottom squarks, i.e., 
$|\sin2\theta_L|$ or $|\sin2\theta_R|$ not far below 1. For the
left-handed squarks this condition poses an obstacle for 
model-building due to the requirement that the CKM mixing-angle 
hierarchy must be reproduced. However, a large mixing between the 
right-handed strange and bottom squarks poses no such problem. In 
fact, several SUSY models of flavor utilizing horizontal 
symmetries exist in which this situation is realized 
\cite{nirseiberg,carone}. 
It therefore appears unlikely that SUSY contributions to the 
SM operators could be as large as those to the 
opposite-chirality operators. Nevertheless, to be fully general 
we will present numerical results for large left-handed or 
right-handed mixing. Finally, as mentioned at the outset, 
significant contributions to the isospin-violating operators
require a large mass splitting between the up and down squarks
of the first generation. This is possible only in the right-handed
sector, since SU(2)$_L$ invariance implies that 
$m_{\tilde d_L}^2=m_{\tilde u_L}^2$ up to tiny SU(2)$_L$-breaking 
corrections. Therefore, only $c_1^{\rm EW}$, $c_2^{\rm EW}$, 
$\widetilde c_3^{\rm EW}$ and $\widetilde c_4^{\rm EW}$ can 
acquire significant gluino box contributions. The magnitudes of 
these contributions are symmetric under interchange of 
$m_{\tilde u_R}^2$ and $m_{\tilde d_R}^2$. One can consider
$m_{\tilde d_R}^2\gg m_{\tilde u_R}^2$ and small 
$m_{\tilde u_R}^2$, or vice versa.

\FIGURE{
\epsfig{file=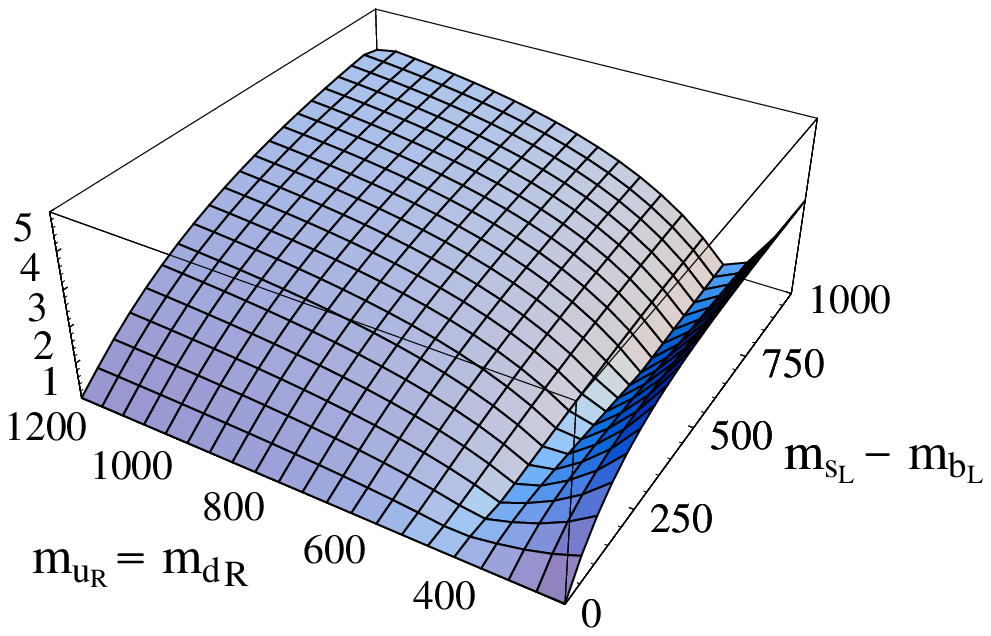,width=7cm}
\hspace{0.2cm}
\epsfig{file=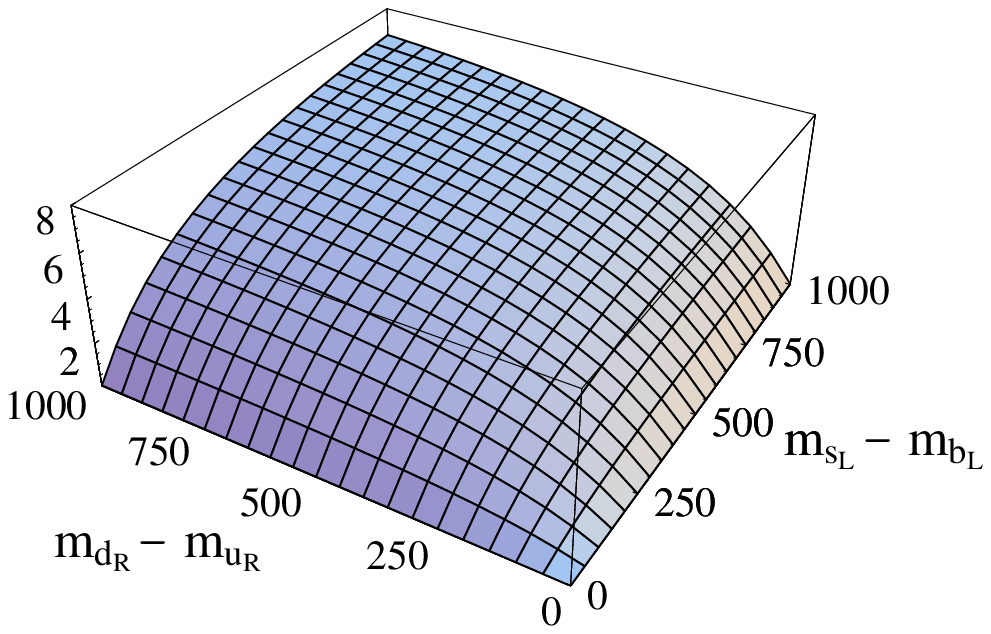,width=7cm} 
\caption{\label{fig:c2}
Left: QCD penguin coefficient $|c_2^{\rm QCD}|$ in units of 
$10^{-4}|\sin2\theta_L|$ versus the common mass $m_{\tilde u_R}
=m_{\tilde d_R}$ (left axis) and the mass splitting 
$m_{\tilde s_L}-m_{\tilde b_L}$ (right axis), for 
$m_{\tilde b_L}=m_{\tilde g}=250$\,GeV. 
Right: Electroweak penguin coefficient $|c_2^{\rm EW}|$ in units 
of $10^{-4}|\sin2\theta_L|$ versus the mass splittings 
$m_{\tilde d_R}-m_{\tilde u_R}$ (left axis) and 
$m_{\tilde s_L}-m_{\tilde b_L}$ (right axis), for 
$m_{\tilde u_R}=m_{\tilde b_L}=m_{\tilde g}=250$\,GeV. All masses 
are given in GeV.}}

We are now ready to present our numerical results. We begin with 
the SUSY contributions to the penguin coefficients at the 
SUSY matching scale, which for simplicity we take to be $m_W$, 
thus ignoring the slow running of $\alpha_s$ and superpartner 
masses above the weak scale. We find that the QCD coefficients 
obey the approximate scaling relation $c_2^{\rm QCD}\sim
c_4^{\rm QCD}\sim -3 c_1^{\rm QCD}\sim -3 c_3^{\rm QCD}$ 
(provided we use the same masses for left- and right-handed 
squarks), which according to (\ref{gluinoQCD}) is exact for the 
contributions of the penguin diagrams but only approximate for the 
box diagrams. The coefficients $\widetilde c_i^{\rm QCD}$ are the 
same as the $c_i^{\rm QCD}$ if all labels $L\leftrightarrow R$ are 
interchanged. In the left-hand plot in Figure~\ref{fig:c2} we show 
the largest coefficient, $c_2^{\rm QCD}$, for a common mass 
$m_{\tilde b_L}=m_{\tilde g}=250$\,GeV as a function of the mass
splitting between the left-handed strange and bottom squarks
and of the common mass $m_{\tilde u_R}=m_{\tilde d_R}$. Note
that only the box contributions in (\ref{gluinobox}) depend on the 
latter two masses, but not the penguin contributions in
(\ref{gluinoQCD}). We find that these two contributions interfere
destructively. For small up- and down-squark masses the box
contributions are dominant, whereas for large masses the boxes 
decouple and the penguin contributions dominate. For intermediate 
masses there is a region with large destructive interference, 
where $c_2^{\rm QCD}$ vanishes or takes small values. Note that 
typical values of the coefficients $c_i^{\rm QCD}$ are of order 
few times $10^{-4}$ (provided the gluino is as light as 250\,GeV 
and there is sufficient mass splitting between the strange and 
bottom squarks), which is an order of magnitude less than the 
typical size of QCD penguin coefficients in the SM \cite{Heff}. If 
the gluino mass is increased and all mass ratios remain the same, 
then the SUSY contributions to the Wilson coefficients decrease, 
scaling like $(250\,\mbox{GeV}/m_{\tilde g})^2$. 

We next turn to the coefficients of the electroweak penguin
operators, which only receive contributions from the gluino box
diagrams. As mentioned above, because of SU(2)$_L$ symmetry only
$c_1^{\rm EW}$, $c_2^{\rm EW}$, $\widetilde c_3^{\rm EW}$ and 
$\widetilde c_4^{\rm EW}$ are important, and we find that for
equal strange--bottom mixing and mass splitting in the 
left-handed and right-handed squark sectors they roughly scale 
according to $c_2^{\rm EW}\sim \widetilde c_4^{\rm EW}\sim 
-3 c_1^{\rm EW}\sim -3\widetilde c_3^{\rm EW}$. In the right-hand
plot in Figure~\ref{fig:c2} we show the largest coefficient, 
$c_2^{\rm EW}$, as a function of the mass splittings between the 
left-handed strange and bottom squarks and between the right-handed
up and down squarks. Provided both splittings are significant, 
the typical values of the electroweak penguin coefficients are of
order few times $10^{-4}$, which is comparable with the value of
the coefficient $c_3^{\rm EW}$ in the SM, given in (\ref{SMinit}). 
Therefore, in certain regions of SUSY parameter space there can
be important isospin-violating contributions to the parameters
$a$ and $b$ entering the $B^\pm\to\pi K$ decay amplitudes.
An important point to notice is that SUSY contributions to the 
electroweak penguin coefficients are typically of same order as, 
or can be larger than, the contributions to the QCD penguin 
coefficients, if there is sufficient mass splitting between the 
right-handed up and down squarks.

\EPSFIGURE{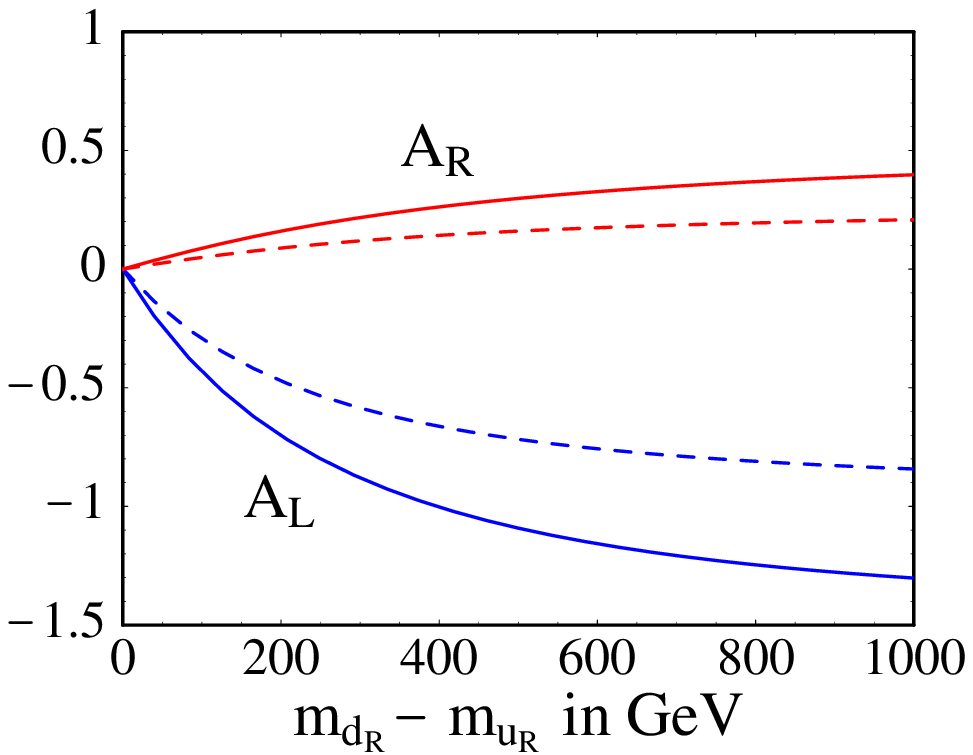,width=7.2cm}
{\label{fig:abSUSY}
Quantities $A_R$ (two upper curves) and $A_L$ (two lower curves) 
versus the mass splitting $m_{\tilde d_R}-m_{\tilde u_R}$, for 
$m_{\tilde u_R}=m_{\tilde b_{L,R}}=m_{\tilde g}=250$\,GeV. The 
solid curves refer to $m_{\tilde s_{L,R}}=1000$\,GeV, the dashed 
ones to $m_{\tilde s_{L,R}}=500$\,GeV.} 

The SUSY contributions to the parameters $a$ and $b$ can be 
decomposed as
\begin{eqnarray}\label{ALAR}
   &&(a_{\rm NP}+ib)_{\rm SUSY} \\
   &&= A_L\sin2\theta_L\,e^{i\delta_L} 
    + A_R\sin2\theta_R\,e^{i\delta_R} \,, \nonumber
\end{eqnarray}
where $A_L$ receives contributions from the electroweak penguin
coefficients $c_1^{\rm EW}$ and $c_2^{\rm EW}$, and $A_R$ 
receives contributions from $\widetilde c_3^{\rm EW}$ and 
$\widetilde c_4^{\rm EW}$. As previously mentioned, to obtain 
large values of these parameters requires a significant mass 
splitting between the right-handed up and down squarks, as well 
as a substantial splitting between the left- or right-handed 
strange and bottom squarks. Exchanging strange and bottom 
or up and down squarks leaves the results invariant up 
to a sign. In Figure~\ref{fig:abSUSY} we show the values of
$A_L$ and $A_R$ versus $m_{\tilde d_R}-m_{\tilde u_R}$ for
two choices of the strange--bottom splitting, such that 
$m_{\tilde s_{L,R}}/m_{\tilde b_{L,R}}=2$ or 4. We note that
for large mass splittings the magnitude of $A_L$ can be up
to twice the SM parameter $\delta_{\rm EW}\approx 0.64$, whereas
for more moderate splittings $|A_L|\sim\delta_{\rm EW}$ can be
obtained. The magnitude of the parameter $A_R$ is typically
smaller by a factor of 3. According to Figure~\ref{fig:shifts}, 
SUSY contributions of this size could lead to shifts of up to 
$\pm 50^\circ$ in the extracted value of $\gamma_{\pi K}$. Even 
in the more realistic case where only right-handed strange--bottom 
squark mixing is large, shifts of up to $\pm 25^\circ$ are 
possible.

\EPSFIGURE{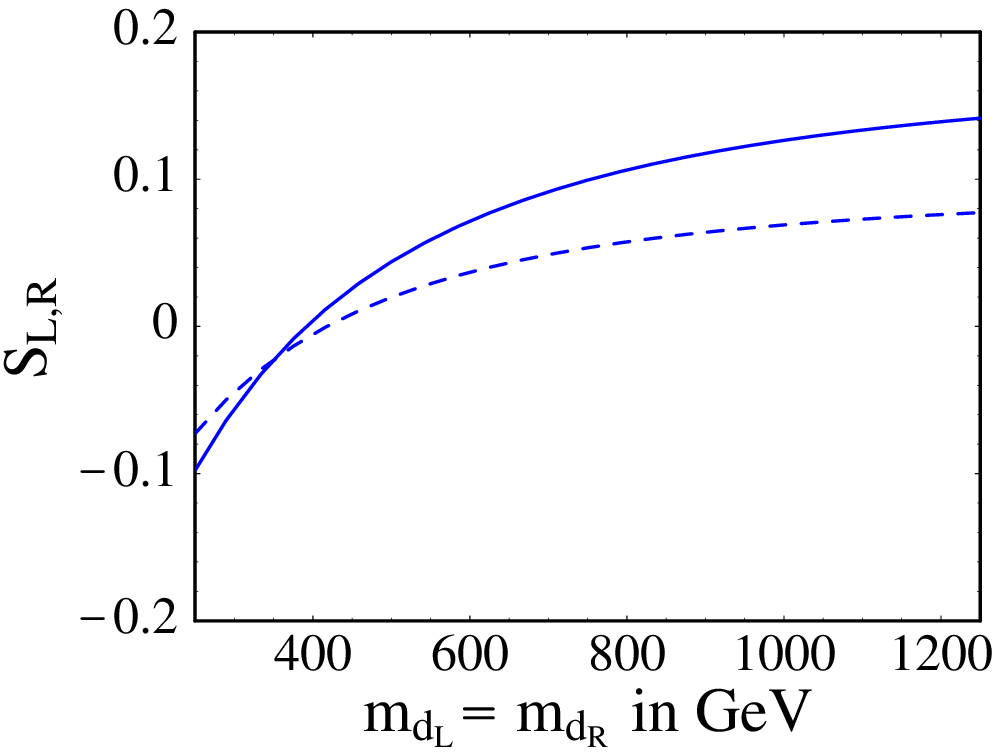,width=7.2cm}
{\label{fig:rhoSUSY}
Quantities $S_{L,R}$ versus the common mass $m_{\tilde d_L}
=m_{\tilde d_R}$, for $m_{\tilde b_{L,R}}=m_{\tilde g}=250$\,GeV. 
The solid curve refers to $m_{\tilde s_{L,R}}=1000$\,GeV, the 
dashed one to $m_{\tilde s_{L,R}}=500$\,GeV.} 

Let us now briefly discuss SUSY contributions to the parameter 
$\rho$ describing CP-violating but isospin-conserving New Physics 
effects in $B^\pm\to\pi K$ decays. From (\ref{rhobound}) it 
follows that QCD penguin coefficients of order few times $10^{-4}$ 
can only lead to rather small values $|\rho|\lsim 0.1$. In analogy
with (\ref{ALAR}), we define
\begin{eqnarray}
   &&(\sin\varphi)_{\rm SUSY} \\
   &&= S_L\sin2\theta_L\sin\delta_L 
    - S_R\sin2\theta_R\sin\delta_R \,, \nonumber
\end{eqnarray}
where $S_L$ ($S_R$) depends on the mass splitting between the 
left-handed (right-han\-ded) strange and bottom squarks, and both 
quantities depend on the masses of the left- and right-handed down 
squarks. In Figure~\ref{fig:rhoSUSY} we show the values of these 
quantities versus the common down-squark mass for two choices of 
the strange--bottom splitting. We see that, indeed, typical 
values of $S_L$ and $S_R$ are of order 0.1 or less. The 
corresponding contributions to $\rho$ are of the same order for
maximal weak phases and large strange--bottom mixing. According to 
Figure~\ref{fig:shifts}, SUSY contributions of this size can only 
lead to insignificant shifts in the extracted value of 
$\gamma_{\pi K}$. 

Thus far we have only considered contributions to $\rho$ due to 
the four-quark penguin operators. The largest possible 
contributions in fact arise if the coefficient $C_{8g}$ of the 
chromomagnetic dipole operator, or the coefficient 
$\widetilde C_{8g}$ of the corresponding dipole operator with 
opposite chirality, have very large magnitudes compared with 
their values in the SM, implying enhanced $\bar b\to\bar sg$ 
transitions. This scenario has been discussed in the context of 
the low semileptonic branching ratio and charm yield in $B$ decays 
\cite{George,Alex,kaon} and the large $B\to X_s\,\eta'$ branching 
ratio \cite{houtseng,SoniA,Alexagain}. In SUSY models it is most 
easily realized via gluino--squark loops containing left--right 
strange--bottom squark mass insertions 
\cite{glue1,Alex,CGGi,berto91}. Constraints on these graphs 
from $B\to X_s\,\gamma$ decays allow for 
$\mbox{B}(B\to X_{sg})\lsim 10\%$, which corresponds to 
$(|C_{8g}|^2+|\widetilde C_{8g}|^2)^{1/2}\approx 1$ (at the 
scale $m_b$), together with possibly large CP-violating phases 
in these coefficients \cite{neub}. From (\ref{rhobound}) it 
follows that large values $\rho=O(1)$ can be obtained in such a
scenario. According to Figure~\ref{fig:shifts}, in this extreme 
case large shifts in the value of $\gamma_{\pi K}$ caused by 
isospin-conserving New Physics are not excluded.

At present, there are no significant phenomenological constraints 
on the angles $\theta_L$ and $\theta_R$ parameterizing the mixing
between the strange and bottom squarks. In particular, the 
measured $B\to X_s\,\gamma$ branching ratio does not impose a 
useful constrain on these parameters. However, an important 
constraint would emerge if in the future $B_s$--$\bar B_s$ mixing 
were found to be consistent with, or not much larger than, its 
predicted value in the SM. For simplicity, we assume that only 
one of the two mixing angles is large. In the case of left-handed
squark mixing, for instance, the relevant gluino box contribution 
to the $B_s$--$\bar B_s$ mass difference $\Delta m_s$, normalized 
to the SM contribution, is given by \cite{berto91}
\begin{eqnarray}
   \left| \frac{\Delta m_s^{LL}}{\Delta m_s^{\rm SM}} \right| 
   &=& \frac{\sin^2\!2\theta_L}{|\lambda_t|^2}\,
    \frac{\alpha_s^2}{\alpha_W^2}\,\frac{m_W^2}{m_{\tilde g}^2}\,
    \frac{1}{C(x_t)}
    \nonumber\\
   &\times& \Bigg| \frac{11}{18} \Big[
    G(x_{\tilde b_L\tilde g},x_{\tilde b_L\tilde g})
    + G(x_{\tilde s_L\tilde g},x_{\tilde s_L\tilde g})
    - 2 G(x_{\tilde b_L\tilde g},x_{\tilde s_L\tilde g}) \Big]
    \nonumber\\
   &&\mbox{}- \frac{2}{9} \Big[
    F(x_{\tilde b_L\tilde g},x_{\tilde b_L\tilde g})
    + F(x_{\tilde s_L\tilde g},x_{\tilde s_L\tilde g})
    - 2 F(x_{\tilde b_L\tilde g},x_{\tilde s_L\tilde g}) \Big]
    \Bigg| \,,
\end{eqnarray}
where $x_t=m_t^2/m_W^2$, and 
\begin{equation}
   C(x) = \frac{x^4-12x^3+15x^2-2x+6x^3\ln x}{4(x-1)^3} \,. 
\end{equation}
For right-handed squark mixing the ratio $|\Delta m_s^{RR}/
\Delta m_s^{\rm SM}|$ would be obtained from the above result via 
the substitution $L\to R$. 

\EPSFIGURE{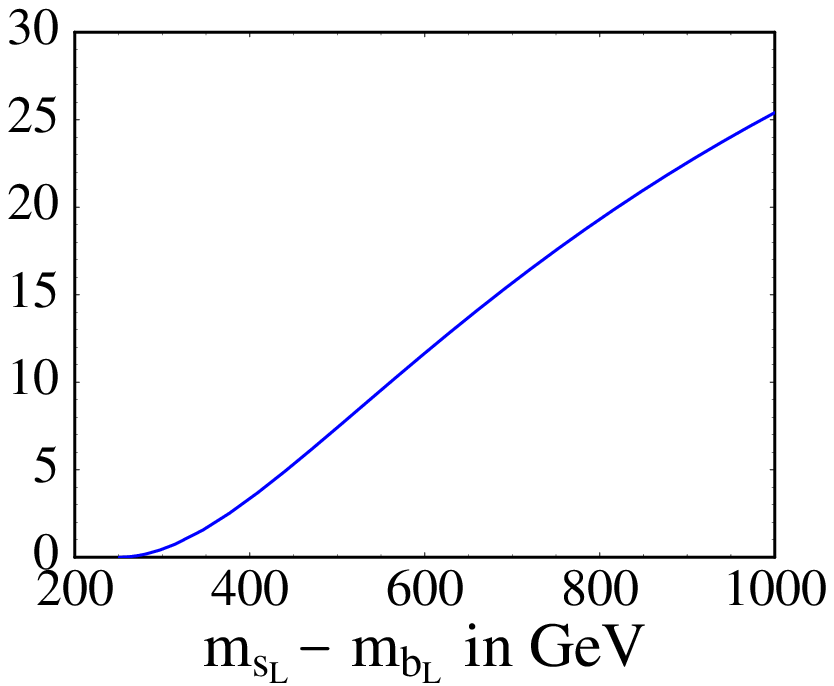,width=7.2cm}
{\label{fig:Bsmix}
Ratio $|\Delta m_s^{LL}/\Delta m_s^{\rm SM}|$ in units of 
$\sin^2\!2\theta_L$ versus the mass splitting 
$m_{\tilde s_L}-m_{\tilde b_L}$ in GeV, for 
$m_{\tilde b_L}=m_{\tilde g}=250$\,GeV.}  

In Figure~\ref{fig:Bsmix} the ratio of the SUSY contribution to
$\Delta m_s$ to the SM result is shown as a function of the mass
splitting between strange and bottom squarks. The same plot with
obvious substitutions $L\to R$ applies to the case of mixing 
between the right-handed squarks. We observe that $\Delta m_s$ 
would greatly exceed the SM value in regions of parameter space 
associated with large SUSY contributions to the penguin 
coefficients. To gauge the potential impact of a $\Delta m_s$ 
measurement near the predicted SM value we impose as an example 
the hypothetical constraint that $|\Delta m_s^{LL}/
\Delta m_s^{\rm SM}|\le 2$. According to Figure~\ref{fig:Bsmix}, 
it then follows that, e.g., $|\sin2\theta_L|<0.52$ for 
$m_{\tilde s_L}=2 m_{\tilde b_L}=500$\,GeV, and 
$|\sin2\theta_L|<0.28$ for $m_{\tilde s_L}=4 m_{\tilde b_L}
=1000$\,GeV. Hence, if such a constraint would have to be imposed 
in the future, the allowed magnitude of the SUSY contributions to 
the penguin coefficients would be reduced by a significant amount. 
(Note, however, that the coefficients of the chromomagnetic dipole 
operators are very weakly constrained by $B_s$--$\bar B_s$ 
mixing.) 

Finally, we comment on implications of naturalness for the large 
right-handed up--down squark mass splitting necessary to obtain 
sizable SUSY contributions to the electroweak penguin 
coefficients. Following \cite{dimo} we note that in models 
in which SUSY is broken at high energies, e.g., supergravity, 
there is a naturalness constraint on the squark and slepton mass 
spectrum coming from the hypercharge $D$-term. In these models the 
$Z$-boson mass is given by
\begin{equation}\label{MZsusy}
   \frac{m_Z^2}{2}
   = \frac{m_1^2-m_2^2\tan^2\!\beta}{\tan^2\!\beta -1} \,,
\end{equation}
where $m_1^2$, $m_2^2$ and $\tan\beta$ are the usual parameters 
of the Higgs potential in the minimal SUSY extension of the SM 
\cite{Hunter}. The renormalization-group equations give
\begin{eqnarray}\label{Dterms}
   m_2^2 &=& -\frac{1-Z_1}{22}\,\mbox{Tr}\,\Big(
    m_{\tilde Q_L}^2 + m_{\tilde d_R}^2
    - 2 m_{\tilde u_R}^2 - m_{\tilde L_L}^2
    + m_{\tilde e_R}^2 \Big) + \hat m_2^2 \,, \nonumber\\
   Z_1 &=& \left( 1 + \frac{33}{20\pi}\,\alpha_{\rm GUT}
    \ln\frac{M_{\rm GUT}^2}{m_{\rm SUSY}^2} \right)^{-1}
    \approx 0.4 \,, 
\end{eqnarray}
where $m_A^2$ ($A=\tilde Q_L, \tilde u_R, \tilde d_R, \tilde L_L, 
\tilde e_R$) are the values of the squark and slepton masses at 
the GUT scale, $\hat m_2^2$ contains the dependence on the other 
soft SUSY breaking masses, and the trace is taken over generation 
space. It is clear from (\ref{Dterms}) that a large mass 
splitting between the first generation right-handed up and down 
squarks poses a potential naturalness problem. However, the 
hypercharge $D$-term can vanish in GUT theories in which 
hypercharge is embedded in the GUT group. For example, in SU(5) 
one has the relations
\begin{equation}
   m_{\tilde Q_L}^2 = m_{\tilde u_R}^2 = m^2_{\tilde e_R}
   \equiv m_{10}^2 \,, \qquad
   m_{\tilde d_R}^2 = m_{\tilde L_L}^2 = m_{\bar 5}^2 \,,
 \end{equation}
so that it is possible to have large up--down squark-mass 
splitting without encountering difficulties with naturalness.  

To summarize, we have seen that SUSY contributions to the Wilson
coefficients of the penguin operators in the effective 
Hamiltonian for $\bar b\to\bar s q\bar q$ transitions, and in
particular of the isospin-violating electroweak penguin operators, 
can be substantial if the gluino and certain squarks have masses
near the weak scale, and other squarks have masses near a TeV. 
Large left-handed or right-handed strange--bottom squark mixing 
is also required. The latter option is naturally realized in 
certain SUSY theories of flavor.

\subsection{Two-Higgs--doublet models}

In extensions of the SM containing charged Higgs bosons 
\cite{Hunter} there are new photon and Z penguin diagrams 
contributing to the Wilson coefficients of the penguin operators. 
Here we consider a general class of two-Higgs--doublet models
(2HDMs) discussed in \cite{Wolf}, which contains the conventional
type-1 and type-2 2HDMs as special cases. We find that, if terms 
of order $m_b/m_t$ are neglected, the new penguin contributions
only involve the $H t_R b_L$ coupling. Following \cite{Wolf} we 
write the corresponding term in the Lagrangian as
\begin{equation}
  {\cal L}_{H t_R b_L} = - \xi_t\,m_t\,
  \bar b_{Lj} V_{ji}^\dagger\,t_{Ri}\,H^- + \mbox{h.c.},
\end{equation}
where $V_{ij}$ is the CKM matrix. In principle, the parameter 
$\xi_t$ may contain a CP-violating phase. However, the penguin 
contributions only depend on $|\xi_t|^2$, and thus any weak 
phase would cancel out. This conclusion holds true even in a 
wider class of multi-Higgs models \cite{YuH}. It is therefore 
sufficient to focus on the conventional type-1 or type-2 
two-Higgs--doublet models, for which $|\xi_t|^2=\cot^2\!\beta$.

It follows from the above discussion that in 2HDMs there are 
no New Physics contributions to the CP-violating parameters
$\rho$ and $b$ entering the parametrization of the 
$B^\pm\to\pi K$ decay amplitudes in (\ref{ampls}). Therefore,
it is sufficient for our purposes to focus on the electroweak
penguin coefficients, which induce a new contribution
to the parameter $a$. Including both the photon and Z penguin 
diagrams, one obtains at the weak scale \cite{1Higgs,MUHiggs}
\begin{eqnarray}
   c_1^{\rm EW} &=& \frac{\alpha\lambda_t}{8\pi} \cot^2\!\beta
    \left[ x_t\,f(x_{tH}) + \frac19\,g(x_{tH}) \right]
    \,, \nonumber\\
   c_3^{\rm EW} &=& \frac{\alpha\lambda_t}{8\pi} \cot^2\!\beta
    \left[ - x_t\cot^2\!\theta_W\,f(x_{tH})
    + \frac19\,g(x_{tH}) \right] \,,
\end{eqnarray}
where $x_{tH}=(m_t/m_{H^+})^2$, and
\begin{eqnarray}
   f(x) &=& \frac{x}{1-x} + \frac{x\ln x}{(1-x)^2} \,,
    \nonumber\\
   g(x) &=& \frac{38x-79x^2+47x^3}{6(1-x)^3}
    + \frac{4x-6x^2+3x^4}{(1-x)^4} \ln x \,.
\end{eqnarray}

In this model there is a simple result for the New Physics
contribution to the parameter $a$, normalized to the SM
contribution. We find
\begin{equation}
   \frac{a_{\rm NP}}{\delta_{\rm EW}}
   = - \cot^2\!\beta\,
   \frac{f(x_{tH})}{1+\frac{3\ln x_t}{x_t-1}}
   \left\{ 1 - 0.74\sin^2\!\theta_W \left[ 1
   + \frac{g(x_{tH})}{9x_t\,f(x_{tH})} \right] \right\} \,,
\end{equation}
where the first term in parenthesis is free of hadronic 
uncertainties, and the second term is numerically small. 

\EPSFIGURE{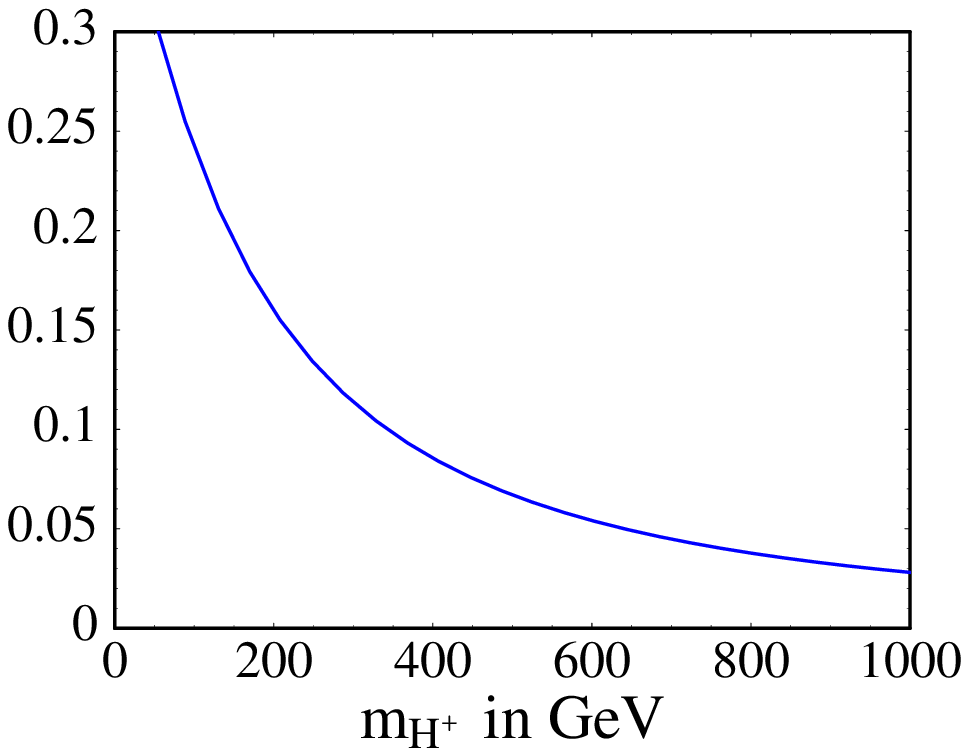,width=7.2cm}
{\label{fig:Higgs}
Ratio $a_{\rm NP}/\delta_{\rm EW}$ in units of $\cot^2\!\beta$ 
in a general 2HDM, as a function of the charged-Higgs mass.}  

In Figure~\ref{fig:Higgs}, we show the ratio 
$a_{\rm NP}/\delta_{\rm EW}$ in units of $\cot^2\!\beta$ as a 
function of the Higgs mass. Even for $\cot\beta=1$ a significant 
contribution to the parameter $a$ requires a small Higgs mass. 
This possibility is not excluded by direct searches,
however in the context of specific models the constraint from 
the $B\to X_s\,\gamma$ branching ratio often favors a larger 
mass and $\cot^2\!\beta<1$ \cite{JoAnne,BarPh}. Therefore, it 
appears unlikely that a large new contribution to $a$ can be 
obtained in 2HDMs. 

\subsection{Models with anomalous gauge-boson couplings}

The SU(2)$_L\times \mbox{U(1)}_Y$ gauge symmetry of the SM 
fully determines the form of the dimension-4 operators that 
describe the vector-boson self-couplings. New Physics may 
induce anomalous couplings of the electroweak gauge bosons,
which at low energies give rise to higher-dimensional operators, 
whose effects are suppressed by inverse powers of the New 
Physics scale $\Lambda$. The effects of anomalous gauge-boson 
couplings on rare $B$ decays have been investigated in 
\cite{bsnn,Gusta}, and their impact on the determination of 
$\gamma$ from $B^\pm\to\pi K$ decays has been discussed in 
\cite{anom}. A general parametrization of the anomalous 
gauge-boson couplings can be found in \cite{HPZK}. In 
low-energy processes, the four new parameters that enter are 
$\Delta\kappa^\gamma$, $\Delta g_1^Z$, $\lambda^\gamma$ and
$g_5^Z$. The first two represent corrections to couplings 
already present in the SM, whereas the latter two refer to new
vertices. Note that these four parameters are real and thus can 
only contribute to the quantity $a$, but not to $\rho$ and $b$.
As in the previous section, we therefore focus only on the 
coefficients of the electroweak penguin operators. They are 
\cite{He-eps}
\begin{eqnarray}
   c_1^{\rm EW} &=& \frac{\alpha\lambda_t}{8\pi}\,x_t \Big[
    \cos^2\!\theta_W\,f_A(x_t) + h_A(x_t) \Big] \,, \nonumber\\
   c_3^{\rm EW} &=& \frac{\alpha\lambda_t}{8\pi}\,x_t \left[
    - \cot^2\!\theta_W \cos^2\!\theta_W\,f_A(x_t)
    + h_A(x_t) \right] \,,
\end{eqnarray}
with
\begin{eqnarray}
   f_A(x) &=& -3\Delta g_1^Z \ln\frac{\Lambda^2}{m_W^2}
    + \frac{6 g_5^Z}{1-x} \left( 1 + \frac{x\ln x}{1-x}
    \right) \,, \nonumber\\
   h_A(x) &=& \frac{\Delta\kappa^\gamma}{2}
    \ln\frac{\Lambda^2}{m_W^2}
    + \lambda^\gamma \left[ \frac{1-3x}{(1-x)^2}
    - \frac{2x^2\ln x}{(1-x)^3} \right] \,, 
\end{eqnarray}
where the New Physics scale $\Lambda$ acts as an ultraviolet 
cutoff. 

As in the case of the 2HDMs, there is a very simple result for 
the New Physics contribution to the parameter $a$, normalized to 
the SM contribution. We find
\begin{equation}
   \frac{a_{\rm NP}}{\delta_{\rm EW}}
   = - \cos^2\!\theta_W\,
   \frac{f_A(x_t)}{1+\frac{3\ln x_t}{x_t-1}}
   \left\{ 1 - 0.74\sin^2\!\theta_W \left[ 1
   + \frac{h_A(x_t)}{\cos^2\!\theta_W\,f_A(x_t)} \right]
   \right\} \,,
\end{equation}
where as before the first term in parenthesis is free of hadronic 
uncertainties. For instance, taking $\Lambda=1$\,TeV gives 
\begin{equation}\label{numbs}
   \frac{a_{\rm NP}}{\delta_{\rm EW}}
   \approx 4.20\Delta g_1^Z - 0.45 g_5^Z
   + 0.19\Delta\kappa^\gamma + 0.03\lambda^\gamma \,.
\end{equation}
From naive dimensional analysis, one expects that 
$\Delta g_1^Z,\Delta\kappa^\gamma\sim(g_W v/\Lambda)^2\sim 
10^{-2}$ (with $v\approx 246$\,GeV the Higgs vacuum expectation 
value), whereas $g_5^Z$ and $\lambda^\gamma$ are expected to 
be further suppressed \cite{Gusta,DaVa}. Potentially the most 
important contribution in (\ref{numbs}) is due to 
$\Delta g_1^Z$, which is bounded by experiment to lie in the
range $-0.113<\Delta g_1^Z<0.126$ \cite{LEP}. Even when
this bound is saturated, $a_{\rm NP}/\delta_{\rm EW}$ cannot 
exceed 0.5 in magnitude. However, from naive dimensional analysis
$a_{\rm NP}$ is naturally an order of magnitude smaller.

\section{Conclusions}
\label{sec:concl}

We have explored how New Physics could affect purely hadronic 
FCNC transitions of the type $\bar b\to\bar s q\bar q$ focusing, 
in particular, on isospin violation. 
Unlike in the Standard Model, where isospin-violating
effects in these processes are strongly suppressed by  
electroweak gauge couplings or small CKM matrix elements, in 
many New Physics scenarios these effects are not parametrically 
suppressed relative to isospin-conserving FCNC processes. 
In the language of effective weak Hamiltonians, this implies
that the Wilson coefficients of QCD and electroweak penguin
operators are of a similar magnitude. For a large class of New 
Physics models, we find that the coefficients of the electroweak 
penguin operators are, in fact, due to ``trojan'' penguins, 
which are neither related to penguin diagrams nor of electroweak 
origin. 

We have calculated the Wilson coefficients of the penguin 
operators in the effective weak Hamiltonian in several New 
Physics models, extending the usual operator
basis where appropriate. We have also included penguin 
operators mediating the decay $\bar b\to\bar d s\bar d$, which
is highly suppressed in the Standard Model. Specifically, we 
have considered: (a) models with tree-level FCNC couplings of 
the $Z$ boson, extended gauge models with an extra $Z'$ boson, 
SUSY models with broken R-parity; (b) SUSY models with R-parity 
conservation; (c) two-Higgs--doublet models, and models with 
anomalous gauge-boson couplings. In case (a), the resulting
electroweak penguin coefficients can be much larger than 
in the Standard Model because they are due to tree-level 
processes. In case (b), these contributions 
can compete with the Standard Model
coefficients because they arise from strong-interaction box
diagrams, which scale relative to the Standard Model like
$(\alpha_s/\alpha)(m_W^2/m_{\rm SUSY}^2)$. In models (c), on 
the other hand, isospin-violating New Physics effects are not
parametrically enhanced and are generally smaller than in
the Standard Model.

We have focused on the rare hadronic decays $B^\pm\to\pi K$,
which are particularly sensitive to isospin-violating effects. 
These decays are especially useful for probing New Physics
contributions, since in the Standard Model the theoretical 
description of such effects is very clean. We have found that 
the ratio $R_*$ of the CP-averaged branching ratios defined in 
(\ref{Rstexp}) and the value of the weak phase $\gamma_{\pi K}$ 
extracted from $B^\pm\to\pi K$ decays are the most useful 
observables for probing isospin-violating New Physics 
contributions. Using a fully general parametrization of the
decay amplitudes, we have derived model-independent bounds
on $R_*$ in the presence of New Physics, and have investigated 
by how much $\gamma_{\pi K}$ could differ from the true value 
of the CKM angle $\gamma$. We have seen 
that, depending on the measured value of $R_*$, it may be 
possible to unambiguously distinguish between isospin-violating 
and isospin-conserving New Physics contributions. Irrespective
of the value of $R_*$, we find that large shifts in $\gamma$ 
can be caused by even moderate isospin-violating contributions 
to the decay amplitudes of order 10\%. In contrast, significant 
shifts due to isospin-conserving New Physics effects would 
require a new $O(1)$ CP-violating contribution to the 
amplitudes.

\TABULAR{|l|cc|cc|}
{\hline\hline
Model & $|a_{\rm NP}+ib|$ & $|\gamma_{\pi K}-\gamma|$
 & $|\rho|$ & $|\gamma_{\pi K}-\gamma|$ \\
\hline
FCNC $Z$ exchange & 2.0 & $180^\circ$ & 0.05 & $3^\circ$ \\
extra $Z'$ boson & 14$^*$ & $180^\circ$ & -- & $180^\circ$ \\
SUSY without R-parity & 14$^*$ & $180^\circ$ & --
 & $180^\circ$ \\
\hline
SUSY with R-parity & 0.4 & $25^\circ$ & 0.12 & $7^\circ$ \\
                   & 1.3 & $180^\circ$ & 0.12 & $7^\circ$ \\
\hline
2HDM & 0.15 & $10^\circ$ & 0 & $0^\circ$ \\
anom.\ gauge-boson coupl.\ & 0.3 & $20^\circ$ & 0
 & $0^\circ$ \\
\hline\hline}
{\label{tab:1}
Maximal contributions to the relevant phenomenological 
parameters, as defined in the text. Entries for $\rho$ 
containing a ``--'' are unconstrained. Entries marked with a 
``$^*$'' are upper bounds obtained using the current values of 
$\bar\varepsilon_{3/2}$ and $R_*$. For the case of SUSY with 
R-parity the first (second) row corresponds to maximal 
right-handed (left-handed) strange--bottom squark mixing. 
For the 2HDM we take $m_{H^+}>100$\,GeV and $\tan\beta>1$.}

For each New Physics model we have explored which regions of 
parameter space can be probed by the $B^\pm\to\pi K$ 
observables, and how big a departure from the Standard Model 
predictions one can expect under realistic circumstances. In 
Table~\ref{tab:1}, we summarize our estimates of the maximal
isospin-violating and isospin-conserving contributions to the
decay amplitudes, as parameterized by $|a_{\rm NP}+ib|$ and 
$|\rho|$, respectively. For comparison, we recall that in
the Standard Model $a\approx 0.64$ and $b\approx\rho\approx 
0$. We also list the corresponding maximal values of the 
difference $|\gamma_{\pi K}-\gamma|$. As noted above, in 
models with tree-level FCNC couplings New Physics effects can 
be dramatic, whereas in SUSY models with R-parity conservation 
isospin-violating loop effects can be competitive with the 
Standard Model. In the case of SUSY models with R-parity 
violation, we have derived interesting bounds
on combinations of the trilinear couplings $\lambda_{ijk}'$
and $\lambda_{ijk}''$, which are given in (\ref{bounds1}) and
(\ref{bounds2}).

It is worth pointing out that isospin- or, more generally, 
SU(3) flavor-violating New Physics effects in hadronic weak 
decays could also be important in other processes. For instance, 
they have been shown to yield a potentially large contribution 
to the quantity $\epsilon'/\epsilon$ 
in $K\to\pi\pi$ decays \cite{epspr}. Moreover, there 
are other $B$ and $B_s$ decay channels that could be sensitive
to flavor-violating New Physics contributions. We look forward 
to returning to this subject in an extra dimension.

\acknowledgments
Y.G.\ and M.N.\ are supported by the Department of Energy under 
contract DE--AC03--76SF00515, and A.K.\ under Grant No.\ 
DE-FG02-84ER40153.

\end{document}